\documentclass[journal, 10pt]{IEEEtran}

\usepackage{amssymb}

\usepackage{amsmath}

\usepackage{subfigure}

\usepackage{times}

\usepackage{graphicx}

\usepackage{url}

\def\bit{BitTorrent}
\def\pex{PEX}

\def\maxoc{maximum number of outgoing connections}
\def\maxps{maximum peer set}
\def\averageps{average peer set}
\def\nated{NATed}
\def\natedlist{L_{nated}}
\def\nonnatedlist{L_{not-nated}}

\def\minneig{minimum number of neighbors before re-contacting the tracker}
\def\returned{number of peers returned by the tracker}
\def\peerset{peer set}
\def\MinNeigh{\delta}

\title{Understanding the Properties of the BitTorrent Overlay}

\specialpapernotice{Technical Report}

\author{\IEEEauthorblockN{Anwar Al-Hamra, Arnaud Legout, and Chadi Barakat\\}
\IEEEauthorblockA{INRIA\\
Sophia Antipolis, France\\
Email: \{anwar.al\_hamra, arnaud.legout, chadi.barakat\}@sophia.inria.fr}}

\begin{document}
\maketitle

\begin{abstract}
  In this paper, we conduct extensive simulations to understand the
  properties of the overlay generated by BitTorrent. We start by analyzing
  how the overlay properties impact the efficiency of BitTorrent. We focus
  on the average peer set size (i.e., average number of neighbors), 
  the time for a peer to reach its
  maximum peer set size, and the diameter of the overlay. In particular, we show
  that the later a peer arrives in a torrent, the longer it takes to
  reach its maximum peer set size. Then, we evaluate the impact of the
  maximum peer set size, the maximum number of outgoing connections
  per peer, and the number of NATed peers on the overlay
  properties. We show that BitTorrent generates a robust overlay, but that
  this overlay is not a random graph.  In particular, the connectivity
  of a peer to its neighbors depends on its arriving order in the
  torrent. We also show that a large number of NATed peers
  significantly compromise the robustness of the overlay to attacks.
  Finally, we evaluate the impact of \textit{peer exchange} on the
  overlay properties, and we show that it generates a chain-like
  overlay with a large diameter, which will adversely impact the efficiency 
  of large torrents. 
\end{abstract}

\section{Introduction}

Recently, Peer-to-Peer (P2P) networks have emerged as an attractive
architecture for content sharing over the Internet.
By leveraging the available resources at the peers, P2P networks 
have the potential to scale to a large
number of peers. Nowadays, P2P networks support a variety of
applications, for instance, file sharing (e.g., \bit{}, Emule), audio
conferencing (e.g., Skype), or video conferencing (e.g., End System
Multicast \cite{chu00case}). Among all existing P2P applications, file
sharing is still the most popular one. A study in 2004 by the
\textit{Digital Music Weblog} magazine \cite{P2P_DigitalMusic} states
that P2P file sharing is responsible for $70-80\%$ of the overall
European Internet traffic. And among the many P2P file sharing
protocols, \bit{} \cite{COHE03_WEP2P} is the most popular one. Alone,
\bit{} generates more than half of the P2P traffic
\cite{P2P_alwaysOn}.

Invented by Bram Cohen, \bit{} \cite{bit_site} targets distributing
efficiently large files, split into multiple pieces, in case of a
massive and sudden demand. The popularity of \bit{} comes from its
efficiency ensured by its peer and piece selection strategies. The
peer selection strategy aims at enforcing the cooperation between
peers while the piece selection strategy tends to maximize the variety
of pieces available among those peers.  The great success of \bit{}
has attracted the curiosity of the research community and several
papers have appeared on this subject. Thanks to this research effort,
we now have a better idea on the strengths and weaknesses of the protocol
\cite{lego06_IMC,lego07_SIGMETRICS,LOCH06_HOTNETS,TIAN06_INFOCOM}.  We
also have a clear idea on the peers' behavior (i.e., arrival and
departure processes), and on the quality of service they experience
\cite{PAM04_BT,GUO05_IMC,QIU04_SIGCOMM}.  
But so far, little effort has been
spent to understand the properties of the distribution overlay
generated by \bit{}.
As already showed by Urvoy et al. \cite{URVO05}, the time to
distribute a file in \bit{} is directly influenced by the overlay
topology. For example, it is reasonable to believe that \bit{}
performs better on a full mesh overlay than on a chain one. In
addition, as compared to a chain, a full mesh overlay makes \bit{}
more robust to peers' departures and overlay partitions.

We conduct in this paper extensive simulations to isolate the main
properties of the overlay generated by \bit{}. Our contributions are
summarized as follows.

\begin{itemize}
\item We first evaluate the impact of the overlay properties on the
  \bit{} efficiency. We show that a large peer set increases the
  efficiency of \bit{}, and that a small diameter is a necessary, but
  not sufficient, condition for this efficiency. We also show
  that the time for a peer to reach its \maxps{} size depends on the size
  of the torrent it joins. The larger the torrent when a peer joins it,
  the longer the time for this peer to reach its maximum peer set size. 

\item We then study the properties of the overlay generated by \bit{}.
  We show that \bit{} generates a graph with with a small diameter. 
  However, this graph is not random and the average peer set size 
  is significantly lower than the maximum possible peer set size. 
  We also show that this overlay is robust to attacks and to churn.

\item We show that the properties of the overlay are not significantly
  impacted by the torrent size, and that a peer set size of 80 is a
  sensible choice. However, a larger peer set size increases the
  efficiency of the protocol at the expense of a higher overhead on
  each peer. We also explain why a maximum number of outgoing
  connections set to half the \maxps{} size is a good choice, and we show
  that a large fraction of \nated{} peers decreases significantly the
  robustness of the overlay to attacks. 

\item Finally, we evaluate the impact of \textit{peer exchange} on the
  overlay properties. Whereas peer exchange allows peers to reach fast
  their maximum peer set size, it builds a chain-like overlay with a large
  diameter. 
\end{itemize}

\newpage{}
The closest work to ours is the one done by Urvoy et al. \cite{URVO05}. 
The authors focus on two parameters, the \maxps{} size and the \maxoc{}. 
As a result, they show that these two parameters influence the distribution 
speed of the content and the properties of the overlay. 

In this paper, we go further and we provide an analysis that highlights the 
relation between the overlay properties and the performance of \bit{}. 
We also present an in-depth study that characterizes the properties of the 
BitTorrent overlay. 
Finally, we show how the overlay properties change as we vary the different system 
parameters. These parameters include, in addition to the \maxps{} size and \maxoc{}, 
the torrent size (i.e., number of peers), the percentage of \nated{} peers, 
and the peer exchange extension protocol. 

The rest of this paper is organized as follows. In Section
\ref{sec:notations} we give a brief overview of \bit{}. In Section
\ref{sec:simulator} we describe our methodology and we give results in
Sections \ref{overlay_impact} and \ref{sec:results}. In Section
\ref{sec:impact_pex} we discuss the impact of peer exchange
on the overlay and we conclude the work with Section \ref{sec:conc}.

\section{Overview of \bit{}}
\label{sec:notations}

\bit{} is a P2P file distribution protocol with a focus on
scalable and efficient content replication. In particular, \bit{}
capitalizes on the upload capacity of each peer in order to increase
the global system capacity as the number of peers increases.  This
section introduces the terminology used in this paper and gives a
short overview of \bit{}.

\subsection{Terminology}
The terminology used in the \bit{} community is not standardized. For
the sake of clarity, we define here the terms used throughout this
paper.

\noindent
\textbf{Torrent:} A torrent is a set of peers cooperating to
  share the same content using the \bit{} protocol.

\noindent
\textbf{Tracker:} The tracker is a central component that stores
  the IP addresses of all peers in the torrent. The tracker is used as
  a rendez-vous point in order to allow new peers to discover existing
  ones. The tracker also maintains statistics on the torrent. Each
  peer periodically (typically every 30 minutes) report, for instance,
  the amount of bytes it has uploaded and downloaded since it
  joined the torrent.

\noindent
\textbf{Leecher and Seed}.  A peer can be in one of two states:
  the \textit{leecher} state, when it is still downloading pieces of
  the content, and the \textit{seed} state, when it has all the pieces
  and is sharing them with others.

\noindent
\textbf{Peer Set:} Each peer maintains a list of other peers to
  which it has open TCP connections.  We call this list the
  \peerset. This is also known as the neighbor set.

\noindent
\textbf{Neighbor:} A neighbor of peer $P$ is a peer in $P$'s peer set.

\noindent
\textbf{Maximum Peer Set Size:} Each peer cannot have a peer set
  larger than the maximum peer set size. This is a configuration
  parameter of the protocol.

\noindent
\textbf{Average Peer Set Size:} The average peer set size is the sum of
  the peer set size of each peer in the torrent divided by the number
  of peers in that torrent.

\noindent
\textbf{Maximum Number of Outgoing Connections:} Each peer has a
  limitation on the number of outgoing connections it can
  establish. This is a configuration parameter of the protocol. 

\noindent
\textbf{Pieces and Blocks:} A file transferred using \bit{} is
  split into pieces, and each piece is split into multiple
  blocks. Blocks are the transmission unit in the network, and peers
  can only share complete pieces with others. A typical piece size is
  equal to 512 kBytes, and the block size is equal to 16 kBytes.

\noindent
\textbf{Official BitTorrent Client:} The official \bit{} client
   \cite{bit_site}, also known as \textit{Mainline} client, was
   initially developed by Bram Cohen and is now maintained by the
   company he founded.

\subsection{ \bit{} Overview}
\label{bit_operations}
Prior to distribution, the content is divided into multiple pieces,
and each piece into multiple blocks. A \textit{metainfo file} is then
created by the content provider.  This metainfo file, also called a
torrent file, contains all the information necessary to download the
content and includes the number of pieces, \mbox{SHA-1} hashes for all the
pieces that are used to verify the integrity of the received data, and
the IP address and port number of the tracker.

To join a torrent, a peer $P$ retrieves the metainfo file out of band,
usually from a well-known website, and contacts the tracker that
responds with an initial peer set of randomly selected peers, possibly
including both seeds and leechers. This initial peer set is augmented
later by peers connecting directly to this new peer. Such peers are
aware of the new peer by receiving its IP address from the tracker.
If ever the peer set size of a peer falls below a given threshold,
it re-contacts the tracker to obtain additional peers.

Once $P$ has received its initial peer set from the tracker, it starts
contacting peers in this set and requesting different pieces of the
content. \bit{} uses specific peer and piece selection strategies to
decide with which peers to reciprocate pieces, and which pieces to ask
to those peers. The piece selection strategy is called the local rarest
first algorithm, and the peer selection strategy is called the choking algorithm. We
describe briefly those strategies in the following. 

\noindent
\textbf{Local rarest first algorithm:} Each peer maintains a list of
the number of copies of each piece that peers in its \peerset{}
have. It uses this information to define a rarest pieces set, which
contains the indices of all the pieces with the least number of
copies. This set is updated every time a neighbor in the peer set
acquires a new piece, and each time a peer joins or leaves the peer
set. The rarest pieces set is consulted for the selection of the next
piece to download.

\noindent
\textbf{Choking algorithm:} A peer uses the choking algorithm to
decide which peers to exchange data with. The choking algorithm is
different when the peer is a leecher or a seed. We only describe here
the choking algorithm for leechers. The algorithm gives
preference to those peers who upload data at high rates.  Once per
\textit{rechoke period}, typically set to ten seconds, a peer
re-calculates the data receiving rates from all peers in its peer
set. It then selects the fastest ones, a fixed number of them, and
uploads only to those for the duration of the period. We say that a
peer unchokes the fastest uploaders via a \textit{regular unchoke},
and chokes all the rest.  In addition, it unchokes a randomly selected
peer via an \textit{optimistic unchoke}. The rational is to discover
the capacity of new peers, and to give a chance to peers with no piece
to start reciprocating.  
Peers that do not contribute should not be able to attain high download rates, 
since such peers will be choked by others. Thus, free-riders, i.e., peers that 
never upload, are penalized. The algorithm does not prevent all free-riding 
\cite{LIOG06_IPTPS,LOCH06_HOTNETS}, but it performs well in a variety of 
circumstances \cite{lego07_SIGMETRICS}. 
Interested readers can refer to Sections 2.3.1 and 2.3.2 in \cite{lego06_IMC} 
for a detailed description of the choking algorithm for leechers and seeds.

\section{Simulation Methodology}
\label{sec:simulator}

To evaluate the properties of the overlay distribution, we have
developed a simulator that captures the evolution of the overlay
over time, as peers join and leave. 
We present here the methodology used and in particular the use 
of simulations over experiments in Section \ref{sim_vs_exp}. 

\subsection{Parameters Used in the Simulations}

\bit{} has the following parameters to adjust the overlay topology:
(1) the \maxps{} size, (2) the \maxoc{}, (3) the \minneig{}, and (4)
the \returned{}. The default value of those four parameters can be
different depending on the version of \bit{}. For example,
the \maxps{} size was recently changed in the mainline client
\cite{bit_site} from $80$ to $200$. Our study shows how these
parameters influence the properties of the overlay and the efficiency
of \bit{}. 

Another parameter that can have an impact on the overlay properties is
the percentage of \nated{} peers. We will evaluate how this parameter 
influences the overlay properties. Note that a \nated{} peer refers to 
a peer behind a NAT or a firewall. 

\subsection{Simulation Details}
\label{simulation_details} 

Our Simulator, that we made public \cite{tracker_simulator_link}, was developed 
in MATLAB. 
We have simulated the tracker protocol as it is implemented in the BitTorrent 
mainline client 4.0.2. In the following, we give the details of our simulator. 
First of all, the tracker keeps two lists of peers, $\natedlist{}$ for \nated{} peers 
and $\nonnatedlist$ for non \nated{} ones. 
Assume that the percentage of \nated{} peers in the torrent is $X\%$. Thus, when a new 
peer $P_i$ joins the torrent, it is considered \nated{} with a probability of $X\%$. 
Then, $P_i$ contacts the tracker, which in turn returns the IP addresses of up to 
$\sigma$ (e.g., $50$) non \nated{} existing peers (if there are any). These $\sigma$ 
IP addresses are selected at random from the $\nonnatedlist$ list. 
Then, the tracker adds $P_i$ to $\natedlist$ if $P_i$ is \nated{} or to $\nonnatedlist$ 
otherwise.

When $P_i$ receives the list of peers from the tracker, it stores them in a list called 
$L_{tracker}^{P_i}$. Then, $P_i$ starts initiating connections to those peers sequentially. 
When $P_i$ initiates a connection to peer $P_j$, $P_i$ removes $P_j$ from $L_{tracker}^{P_i}$. 
When a peer $P_j$ receives a connection request from peer $P_i$, $P_j$ 
will accept this connection only if its peer set size is less than the \maxps{} size. 
In this case, $P_i$ adds $P_j$ to its list of neighbors $L_{neighbors}^{P_i}$. $P_j$ also 
adds $P_i$ to $L_{neighbors}^{P_j}$. 
Note that, in practice, peer $P_i$ would initiate TCP connections to the peers in its 
$L_{tracker}^{P_i}$. In our simulator, establishing a connection between peers $P_i$ 
and $P_j$ results in adding $P_i$ to the list of neighbors of $P_j$ $L_{neighbors}^{P_j}$ 
and also adding $P_j$ to the list of neighbors of $P_i$ $L_{neighbors}^{P_i}$. 
This is reasonable because our goal is to reproduce the topology properties of the overlay 
and no data exchange is simulated over the links between peers.

After the connection has been accepted or refused by $P_j$, $P_i$ initiates a new 
connections to the next peer in $L_{tracker}^{P_i}$. 
Peer $P_i$ keeps on contacting the peers it discovered from the tracker until 
(1) it reaches its \maxoc{}, or (2) $L_{tracker}^{P_i}$ becomes empty.

Assume that $P_i$ and $P_j$ are neighbors. When $P_j$ leaves the torrent, $P_i$ removes $P_j$ 
from its list of neighbors $L_{neighbors}^{P_i}$. The tracker also removes $P_j$ 
from $\natedlist$ or $\nonnatedlist$ depending on whether $P_j$ was \nated{} or not. 
In addition, $P_i$ will try to replace the neighbor it lost. For this purpose, $P_i$ checks 
whether the number of connections it has initiated to its actual neighbors is less than the 
\maxoc{}. If this is the case, $P_i$ checks whether it still knows about other peers in the torrent, 
i.e., if $L_{tracker}^{P_i}$ is not empty. If this is the case, it contacts them sequentially until 
either (1) one of them accepts the initiated connection or (2) $L_{tracker}^{P_i}$ becomes empty. 

Whenever the peer set size of $P_i$ falls below a given threshold (typically $20$), it 
recontacts the tracker and asks for more peers. We set the minimum interval time between 
two requests to the tracker to $300$ simulated seconds. 

Finally, each peer contacts the tracker once every $30$ minutes to indicate that it is still 
present in the network. If no report is received from a peer within $30-45$ minutes, the tracker 
considers that the peer has left and deletes it from $\natedlist$ or $\nonnatedlist$. 

Our simulator mimics the real overlay topology construction in BitTorrent and therefore, 
we believe that our conclusions will hold true for real torrents. 

\subsection{Simulations vs. Experiments}
\label{sim_vs_exp}
There are three reasons that motivated us to perform simulations and not 
experiments. 
First, in BitTorrent, we cannot use solutions that rely on a crawler to infer the topology 
properties as already done in the context of Gnutella \cite{STUT05_SIGMETRICS}. The reasons 
is that 
BitTorrent does not offer distributed mechanisms for peers discovery or data lookup. Thus, 
there is no way to make a BitTorrent peer give information about its neighbors. 

Second, we cannot take advantage of existing traces collected at various trackers. The reason 
is that a peer never sends information to the tracker concerning its connectivity with other 
peers, e.g., its list or number of neighbor. 

Third, we can experimentally create our own controlled torrents. However, in order 
to give significant results, we need torrents of moderate size with more than $1000$ 
peers, which is not easy to obtain. 
In addition, as we have much less flexibility with real experiments, and as we are 
only concerned by the overlay construction (except in Section \ref{discussion_outdegree}) 
which is far easier to simulate than the data exchange protocol, we decided to run 
simulations instead of experiments. We validate our simulator on a small torrent of $420$ 
peers in Section \ref{val_exp}.

\subsection{Arrival Distribution of Peers}
\label{sec:arriv-distr-peers}
 \begin{figure}[tb]
 \centering
 \includegraphics[height=0.28\textwidth]{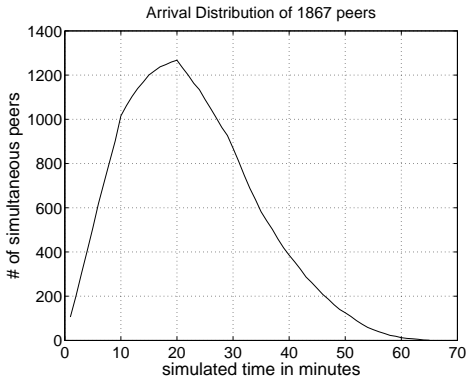}
 \caption{\label{arrival_process} The evolution of our torrent size over time. 
 The overall number of peers that join this torrent from the beginning to the 
 end is equal to $1867$, and the maximum number of simultaneous peers is about 
 $1250$. {\it{The rate at which peers join the torrent decreases exponentially with 
 time.}}}
 \end{figure}

We assume that peers' arrivals follow an exponential distribution,
i.e., the rate at which peers join the torrent decreases 
exponentially with time.  More precisely, we split the simulated 
time into slots. Each slot represents $10$ minutes of simulated time. 
The first slot of time refers to the first $10$ simulated minutes. 
More formally, slot $i$ is defined as the simulated time elapsed 
between the moment $t=(i-1)\cdot 10$ minutes and the 
moment $t=i\cdot 10$ minutes. Then, within each slot of time $i$, 
the number of new peers that join the torrent is computed as 
\begin{eqnarray}
\label{arrival_eq}
arrivals~at~slot~i &=& 1000\cdot \exp^{-0.7\cdot (i-1)} ~if~i \leq 4\\
 		   &=& 0 ~if~ i > 4 \notag
\end{eqnarray}
Each peer stays on-line for a random amount of time uniformly 
distributed between $10$ and $20$ simulated minutes. 
Under this assumption on the arrival process, $1000$ peers will
arrive during the first $10$ minutes of the simulations, $497$ peers during 
the second $10$ minutes, $247$ peers during the third $10$ minutes, $123$ peers 
during the fourth $10$ minutes. Note that no peer will arrive after the first $40$ 
minutes of the simulation. 
As a result, we have more arrivals than departures during the first two time slots. 
In contrast, starting from the third time slot, the departure rate becomes higher 
than the arrival rate. 
The torrent size that results from these arrivals and from the
lifetime distribution described above corresponds to a typical
torrent size evolution \cite{PAM04_BT,GUO05_IMC}. 
The overall number of peers that join this torrent from the beginning
to the end is equal to $1867$, and the maximum number of simultaneous
peers is about $1250$. 

Even if this torrent is of moderate size, we will show later that
it allows us to gain important insights on the
properties of the overlay. Moreover, we will explain how we can
extrapolate our results to larger torrents.
Note that, the lifetime of our torrent is of $70$ simulated minutes and 
the average lifetime of a peer is $15$ minutes. One may wonder whether this 
is realistic as BitTorrent is mostly used to download large files. 
Typically, the lifetime of a BitTorrent's peer is of several hours and the 
torrent's lifetime ranges from several hours to several months. 
However, we are interested only in the construction of the overlay and not in the 
data exchange. Thus, we only need to see how the overlay adapts dynamically to 
the arrival and departure of peers, which is ensured by the arrival distribution 
we consider. 
As a result, considering torrents and peers with larger lifetimes will 
not give any new insights. It will only increase the run time of the simulations. 

\subsection{Metrics}
\label{sec:metrics}
We consider $4$ different metrics to evaluate the overlay properties in
this paper. Those metrics are discussed below.

\noindent
{\bf{Average peer set size:}} The peer set size is critical to the
  efficiency of \bit{}. Indeed, the peer set size impacts the piece 
  and peer selection strategies, which are at the core of the \bit{} 
  efficiency.

  The piece selection strategy aims at creating a high diversity of
  pieces among peers. The rational is to guarantee that each peer can
  always find a piece it needs at any other peer.  This way, the peer
  selection strategy can choose any peer in order to maximize the
  efficiency of the system, without being biased by the piece
  availability on those peers.
  However, this piece selection strategy is based on a version of
  rarest first with local knowledge. Whereas with global rarest first
  each peer replicates pieces that are globally the rarest, with local
  rarest first each peer replicates pieces that are the rarest in its
  peer set. Therefore, the peer set size is critical to the efficiency
  of local rarest first. The larger the peer set, the closer local
  rarest first will be to global rarest first.

  The peer selection strategy aims at encouraging high peer
  reciprocation by favoring peers who contribute. Recently, Legout et
  al. \cite{lego07_SIGMETRICS} showed that peers tend to unchoke more
  frequently other peers with similar upload speeds, since those are
  the peers that can reciprocate with high enough rates. Thus, the
  larger the peer set, the higher the probability that a peer will
  find peers with similar upload capacity and the more efficient the
  choking algorithm. We confirm this analysis with simulations in
  Section~\ref{discussion_outdegree}.

\noindent
 {\bf{Speed to converge to the maximum peer set:}} As we will show later, 
 a large peer set helps the peer to progress fast in the download of
 the file. Thus, it is important to investigate how long a peer takes 
 in order to reach its \maxps{} size.

  \noindent {\bf{Diameter of the overlay distribution:}} A short
  diameter is essential to provide a fast distribution of pieces. In
  Section~\ref{discussion_diameter}, we develop a simple analysis to
  support this claim.

\noindent
{\bf{Robustness of the overlay to attacks and high churn rate:}} 
  P2P networks represent a dynamic environment where peers can join 
  and leave the torrent at any time. As a result, it is important to 
  know whether the overlay generated by \bit{} is robust to high churn 
  rate. In addition, P2P overlays may be subject to attacks that target 
  to partition the overlay. In Section \ref{sec:results}, we explain how 
  we simulate churn rates and attacks.

\section{Impact of the Overlay on \bit{}'s efficiency}
\label{overlay_impact}

In this section, we investigate the impact of the overlay structure on
the efficiency of \bit{}.  First, we evaluate the impact of the peer
set size on \bit{} efficiency. Second, we analyze the convergence
speed of peers toward their \maxps{} size. Finally, we develop a simple
model that highlights the relation between the diameter and the
distribution speed of pieces.
Note that the robustness of the overlay will be studied in
Section~\ref{sec:results} through simulations.

\subsection{Impact of the Average Peer Set Size}
\label{discussion_outdegree}

In this section, we simulate the exchange of pieces in \bit{} in order to understand
the influence of the \averageps{} size on the efficiency of the protocol.
Our simulator runs in rounds where each round corresponds to 10 
simulated seconds, which is the typical duration between two calls of 
the choking algorithm in \bit{}. Every 10 seconds, we scan all peers
one after the other.  For each peer, we apply the choking algorithm to
identify the set of peers it is actively exchanging data with. Then,
we apply the piece selection strategy to discover which pieces to
upload to each peer chosen by the choking algorithm. The choking algorithm is
implemented as explained in Section~2.3.2 in \cite{lego06_IMC}.  We
consider that bottlenecks are at the access links of the peers. We do not
consider network congestion, propagation delays, and network failures.

We generate three overlays each with $1000$ peers and a diameter of $2$. They 
only differ in their peer set size. The first overlay has a peer set size of 
$50$, the second one has a peer set size of $100$, and the third one has a peer 
set of $150$. 
We now explain how to construct an overlay with $1000$ peers, 
with a diameter of $2$ and a peer set size of $50$. The same 
methodology is used to construct the two other overlays.
We apply this algorithm for each peer sequentially starting with 
$P_1$. For each peer $P_i$, we connect it to other 
peers randomly selected from the set of peers $\{P_1, \dots, P_{1000}\}$. The 
following two conditions should never be violated. First, no peer is allowed to 
have more than $50$ neighbors. Second, a peer cannot be its own neighbor. 
Note that there is no guarantee that each node will have exactly $50$ neighbors. 
Yet, our results show that very few peers have less than $50$ neighbors. Then, we 
select one source at random to distribute a file of $100$MB, which is
split into $100$ pieces. We assume that all peers join the torrent at
its beginning and stay until the end of the
simulation. Each peer has a download capacity of 1 Mbit/s, and the
upload capacity is randomly selected, with a uniform distribution,
between $160$ kbit/s and $350$ kbit/s. 
This homogeneous download capacity of peers is reasonable as the replication 
speed of pieces is usually limited by the upload capacity of peers.  
Our goal is to study the
evolution of the total number of pieces received by each peer in the
torrent.

\begin{figure}[tb]
\centering
\includegraphics[height=0.28\textwidth]{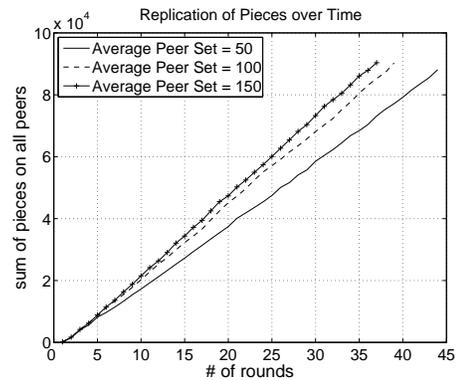}
\caption{\label{impact_outdegree}  Impact of the peer set size on the \bit{} 
efficiency. These results are the average over ten independent runs. 
{\it The larger the peer set size, the faster the replication of pieces.}}
\end{figure}

Our results confirm previous ones showed by Urvoy et al. \cite{URVO05} and 
show that \bit{} replicates pieces faster with a larger peer set. Indeed, Fig. 
\ref{impact_outdegree} shows that, as we increase the peer set size 
from $50$ to $100$ (respectively from $100$ to $150$), the replication speed 
of pieces improves by $12\%$ (respectively by $5\%$). 

In summary, a larger peer set improves the speed of piece
replication. However, this is at the expense of an additional load on
each peer that has to maintain a larger number of TCP connections and
has to handle an additional signaling overhead per connection. 
Keep in mind that, in the following, and while evaluating the overlay 
properties, there will be no data exchange between peers. We now only focus 
on the evolution of the overlay as peers join and leave the torrent. 

\subsection{Analysis of the Convergence Speed}
\label{discussion_convergence}
A \bit{} client usually needs time to reach its \maxps{} size. In this
section, we show that this is a structural problem in \bit{} and that
the convergence speed depends on the torrent size and on the arrival rate
of peers. We consider in our analysis a \maxps{} size of $80$, a \maxoc{}
of $40$, and $50$ peers returned by the tracker. We have chosen fixed
values for the sake of clarity, and it is straight forward to extend our
analysis to other parameter sets.

When a new peer $P_i$ joins the torrent, it receives from the tracker
the IP addresses of 50 peers chosen at random among all peers in the
torrent. Then, $P_i$ connects to at most 40 out of these 50 peers. To
complete its peer set and have $80$ neighbors, $P_i$ keeps on
cumulating new connections received from the peers that arrive after
it. One can easily derive on average how long a peer needs to wait
until it completes its peer set.

\begin{figure}[tb]
\centering
\includegraphics[height=0.28\textwidth]{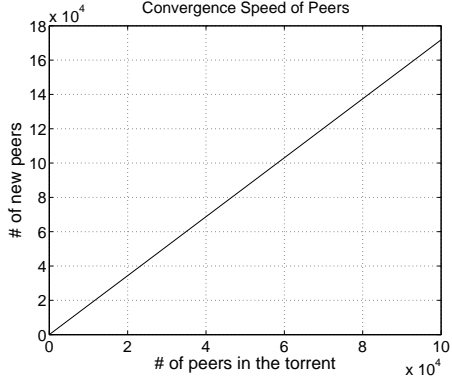}
\caption{\label{fig:convergence_time} Convergence speed of peers
  toward their \maxps{} size. {\it The number of peers that should arrive
    after peer $P_i$ so that $P_i$ receives $40$ incoming connections
  depends linearly on the size of the torrent when $P_i$ has joined
  the torrent.}}
\end{figure}

We assume that the number of peers in the torrent is $N_s$ when $P_i$
arrives. We also assume that $P_i$ has succeeded to initiate $40$
outgoing connections and still misses $40$ incoming connections in
order to reach its maximum peer set of $80$ connections. Therefore,
the probability that peer $P_i$ receives a new connection from a new
peer $P_j$ joining the torrent is $\frac{40}{N_s}$. Thus, the number
of peers $K$ that should arrive after peer $P_i$ in order for this
peer to cumulate $40$ incoming connections is on average given by:
\begin{equation}
\label{eq:convergence_time}
1 = \sum_{n = N_s+1}^{N_s+K} \frac{1}{n}
\end{equation}
Fig. \ref{fig:convergence_time} shows $K$ as a function of $N_s$ as
obtained from Eq. (\ref{eq:convergence_time}). We see that the time
for a peer to reach its maximum peer set size increases linearly with the
torrent size $N_s$. For example, peer $P_{100}$ should wait the arrival 
of $173$ peers after it in order to receive $40$ incoming connections, 
peer $P_{1000}$ should wait for $1720$ peers and peer peer $P_{10000}$ 
for $17184$ peers. 

This linear dependency can be further shown through the following approximation 
obtained from Eq. \ref{eq:convergence_time_approx}:
\begin{equation}
\label{eq:convergence_time_approx}
K \sim (e - 1)\cdot N_a 
\end{equation}
The error generated by this approximation is low even for very small 
torrent sizes. For example, we obtain an error of $0.6\%$ for a torrent of $100$ 
peers and an error of $0.09\%$ for a torrent of $1000$ peers.

In summary, the larger the torrent when a peer joins it, the longer
this peer will wait to reach its \maxps{} size. 

\subsection{Impact of the Diameter of the Distribution Overlay}
\label{discussion_diameter}

Yang et al. \cite{YANG04_p2p} shows that the service capacity of
P2P protocols scales exponentially with the number of peers
in the torrent. In this section, we apply their analysis to show the
impact of the diameter on the capacity of service of P2P
protocols.

We consider a torrent with $N=2^k$ peers. We assume that all peers
have the same upload and download capacity $b$. Moreover, we assume
that all peers join the system at time $t=0$ and stay until the file
is distributed to all peers. The unit of time is $T=\frac{C_s}{b}$, where
$C_s$ is the content size.
Each peer downloads the content from a single peer at a time. A peer can start
uploading when it receives entirely the content. Then, it can upload to
a single peer at a time.
We finally assume that the file is initially available only at the source $S$.

At time $t=0$, $S$ starts serving the file to peer $P_1$. At time
$t=T$, $P_1$ receives completely the file and starts serving it to
peer $P_2$. At the same time, $S$ schedules a new copy of the file to
peer $P_3$. After $l.T$ units of time, $2^l$ peers have entirely
the file. As a result, the number of sources, thus the capacity of
service, scales exponentially with time.
However, this means that, at any time $i\cdot T$, the $2^i$ sources of the 
file should find $2^i$
other peers that have not yet received the content. In other words, each
of the $2^i$ sources must have a direct connection in the overlay to a
different peer that does not have yet the content.
Whereas it is likely that such a condition is verified most of the
time for an overlay with a small diameter, it is not clear what
happens when the diameter is large.

\begin{figure}[tb]
\centering
\includegraphics[height=0.20\textwidth]{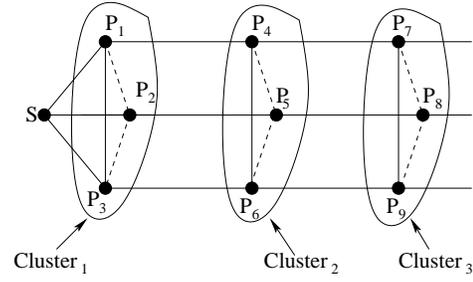}
\caption{\label{chain_graph} An example of a chain-like overlay.
For clarity, we consider only 3 chains that expand in parallel. At
each level in this overlay, we have a cluster that includes three
peers fully interconnected with each other.
\it{The distribution time of the file increases linearly with the number of clusters in the overlay.}}
\end{figure}

Now, we evaluate how the capacity of service scales on the chain-like
overlay shown in Fig. \ref{chain_graph}.  The overlay includes
multiple levels. At each level, we have a cluster that includes $2^m$
peers. Each peer is connected to all peers in its cluster. In
addition, it maintains one connection to the two clusters that
surround its own cluster. The source is connected only to the peers in
the first cluster.
For this overlay, at time $t=0$, the source serves the file to peer
$P_1$ in the first cluster. At time $t=T$, after receiving the
entire file, $P_1$ starts serving it to the peer it knows in the
second cluster while $S$ schedules a new copy of the file to peer
$P_2$, in the first cluster. At time $t=2.T$, $P_1$ does not know any
other peer in the second cluster and therefore, it will serve the file
to a new peer in its own cluster. We can easily verify that the
intra-cluster capacity of service, i.e., the capacity of service
inside a cluster once it has at least once source of the content,
increases exponentially with time. However, the inter-cluster capacity
of service, i.e., the time for each cluster to have at least one
source of the content, increases linearly with time. In other words,
once the file is served by the source $S$, it needs $p.T$ units of
time to reach the cluster $P$. Thus, for a chain with $N_{clusters}$ of
size $2^m$ each, the service time of the file is $(N_{clusters} + m).T$,
where $N_{clusters}.T$ units of time are needed to reach the $N^{th}$
cluster in the overlay and $m.T$ units of time are needed to duplicate
the file over the $2^m$ peers inside the last cluster.
As a result, a chain-like overlay fails to keep peers busy all the
time and, consequently, the distribution time of the file increases
linearly with the number of clusters in the system.

In summary, a short diameter is necessary to have a capacity of
service that scales exponentially with time. However, this condition
is not sufficient. For instance, a star overlay where all
peers are connected to the same peer in the center has a diameter of
2. In this topology, the peer at the center of the star is a
bottleneck, which leads to a poor capacity of service. Therefore, when
analyzing the properties of an overlay, the diameter should be
interpreted along with the shape of the overlay before making any
conclusion.

\section{Characterizing the Properties of \bit{}'s Overlay}
\label{sec:results}

In Section~\ref{overlay_impact} we have evaluated the impact of the
overlay properties on \bit{} efficiency. In this section, we conduct
extensive simulations in order to study the properties of the overlay
generated by \bit{}. We first describe the properties of the overlay
for a default set of parameters. Then, we vary some of them in order to
identify their impact on the overlay properties. For each simulation,
we evaluate the overlay properties according to the four metrics
introduced in Section~\ref{sec:metrics}. 

\subsection{Initial Scenario}
\label{basic_scenario}
For this initial scenario, we consider a set of parameters used by
default in several \bit{} clients (e.g., mainline 4.x.y  \cite{bit_site}).  We set the
\maxps{} size to $80$, the \maxoc{} to $40$, the number of peers returned
by the tracker to $50$, and the minimum number of peers to $20$. We
also consider a torrent with $1867$ peers, as described in
Section~\ref{sec:arriv-distr-peers}, and we assume that no peer is
\nated{}. We study the case of \nated{} peers in
Section~\ref{sec:impact-number-nated}. 

\begin{figure}[tb]
\centering
\includegraphics[height=0.28\textwidth]{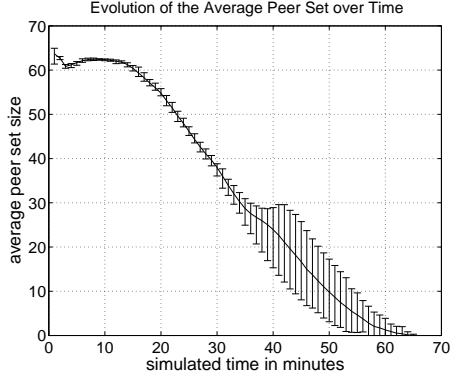}
\caption{\label{tracker_outdegree} The \averageps{} size over time for
  the initial scenario, averaged over ten independent runs. 
  At a given time $t$, the average peer set size is the sum of the peer set 
  size at time $t$ of each peer in the torrent divided by the number of peers 
  in that torrent.
  The error bars indicate the minimum and maximum. 
  {\it The \averageps{} size is lower than the \maxps{} size.}}
\end{figure}

\textbf{Average peer set size:} As we can see in Fig.
\ref{tracker_outdegree}, \bit{} generates an \averageps{} size that is
low compared to the \maxps{} size targeted. For example, the
\averageps{} size does not exceed 65 while the \maxps{} size is set to
80.  To explain this low \averageps{} size we focus on the convergence
speed.

\begin{figure}[tb]
\centering
\includegraphics[height=0.28\textwidth]{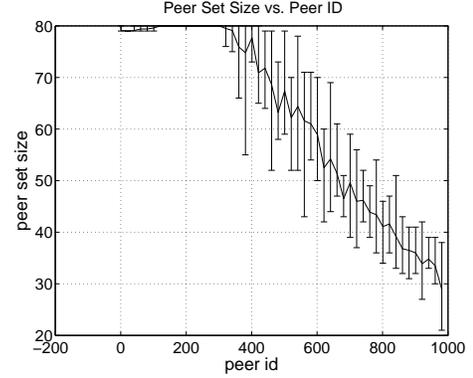}
\caption{\label{outdegree_distribution} The peer set size as a function of the 
  peer id at time $t=10$ minutes, averaged over ten independent runs. 
  The error bars indicate the minimum and maximum. 
  Peers are ordered according to
  their arriving time, the smaller the index the earlier the arrival
  time. {\it The later a peer joins the torrent, the smaller its peer
    set size.}}
\end{figure}

\textbf{Convergence speed:} In Section~\ref{discussion_convergence}, we
have seen that the time for a peer to reach its maximum peer set size
depends on the torrent size at the moment of its arrival and on the
arrival rate of new peers. In Fig. \ref{outdegree_distribution}, we
depict the distribution of the \peerset{} size after $10$ simulated 
minutes over
the first $1000$ peers in the torrent.  As we can see, the later a
peer joins the torrent, the smaller its peer set size. More precisely,
the peers that join the torrent earlier reach their \maxps{} size. In
contrast, the peers that arrive later do not cumulate enough incoming
connections to saturate their peer set, and therefore, their peer set
size is around the \maxoc{} of $40$. As a result, the \averageps{}
size is only $65$.
In Fig. \ref{outdegree_distribution}, one would expect peer $P_{1000}$
to have a peer set size of $40$ and not $30$.  In fact, among the $50$
peers returned by the tracker at random, only $30$ had a peer set size 
lower than 80. This is also the reason behind the oscillations in this
figure.

\begin{figure}[tb]
\centering
\includegraphics[height=0.28\textwidth]{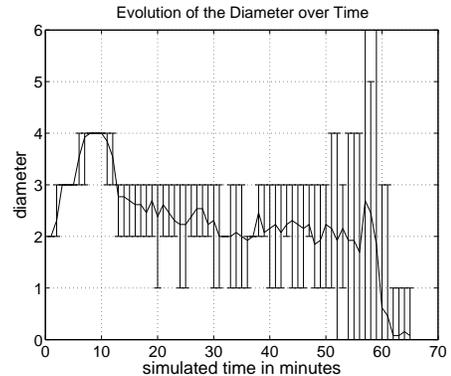}
\caption{\label{tracker_diameter} The diameter of the overlay over
  time for the initial scenario, averaged over ten independent runs. 
  The error bars indicate the minimum and maximum. 
{\it The average diameter is lower than 4 during the entire simulations.}}
\end{figure}
\textbf{Diameter of the overlay:} \bit{} generates an overlay with a
short diameter. As we can observe in Fig.  \ref{tracker_diameter}, the
average diameter of the overlay is between $2$ and $4$ most of the 
time\footnote{In this paper, we compute the diameter of the overlay as the longest shortest path between $1000$ peers of the overlay selected at random. This method allows us to obtain very good approximation of the diameter and speeds up the run time of the simulator.}.
However, at the end time of the torrent, the overlay may get partitioned. 
If we look closely at the right part of Fig. \ref{tracker_diameter}, we 
notice that the minimum value of the diameter goes to zero, a value we use 
to indicate partitions. Actually, after a
massive departure of peers, we may obtain many small partitions each
of tens of peers. The partitions are due to the minimum number of
neighbors $\MinNeigh{}$ a peer should reach before recontacting the
tracker for new peers. Indeed, to minimize the interaction between
peers and the tracker, a peer asks the tracker for more peers only if
its number of neighbors falls below $\MinNeigh$. Therefore, as long as
the peer set size is larger than $\MinNeigh$, to recover from a
decrease in its peer set size, a peer has to wait for new incoming
connections from newly arriving peers, which does not happen toward
the end of the torrent. That is why the torrent does not merge
again. To prevent such a behavior, one needs to assign a high value to
$\MinNeigh{}$. The value of $\MinNeigh{}$ is then a trade-off between
having a connected overlay at the end of the torrent, and a high load
at the tracker.

\begin{figure}[tb]
\centering
\includegraphics[height=0.28\textwidth]{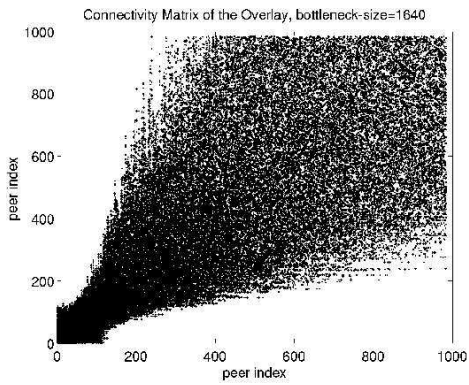}
\caption{\label{connectivity_event10}The connectivity matrix of the
  \bit{}'s overlay after 10 minutes, for a single run. We only show the first $1000$
  peers. A dot at (i,j) means that i and j are neighbors. 
  We have a bottleneck of $1640$, which refers to the number of connections between 
  the first $80$ peers and the rest of the torrent.
  {\it \bit{} does not generate a random graph.}}
\end{figure}

In our analysis in Section~\ref{discussion_diameter}, we show that
a short diameter is a necessary, but not sufficient, condition for an
efficient distribution of the file. We
draw in Fig. \ref{connectivity_event10} the shape of the overlay
generated by \bit{}. 
We see that \bit{} does not generate a random overlay, and that the
overlay has a specific geometry. 
Indeed, Fig. \ref{connectivity_event10} shows a clustering among peers that
arrive first. For example,
peer $P_{25}$ is connected only to the first hundred peers. The
reason is that when $P_{25}$ arrives, it connects to all the $24$
peers already existing in the torrent. Then, $P_{25}$ waits for new
arrivals in order to complete the 56 peers it still needs to
saturate its peer set. According to Eq. (\ref{eq:convergence_time}),
these missing connections can be fulfilled after the arrival of 75
peers on average, $P_{26}, \dots, P_{100}$. Similarly, when peer
$P_{200}$ arrives, it establishes up to 40 outgoing connections.
However, $P_{200}$ needs to wait the arrival of a large number of
peers in order to complete its 40 incoming connections. This explains
why, as compared to $P_{25}$, the neighbors of $P_{200}$ are
selected from a larger set of peers (i.e., between $P_{60}$ and
$P_{600}$). 
Even though we have this clustering phenomena, the overlay does not 
include bottlenecks. Actually, the number of connections between 
the first $80$ peers and the rest of the network is equal to $1640$. 
Therefore, the \bit{} has the potential to allow a fast expansion of 
the pieces. 

\begin{figure}[tb]
\centering
\includegraphics[height=0.28\textwidth]{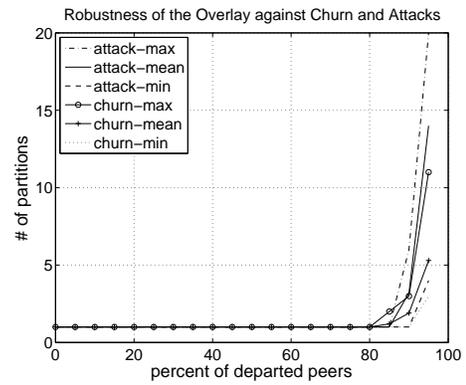}
\caption{\label{robustness_basic} The number of partitions in the
  overlay as a function of the percentage of departed peers, averaged
  over ten independent runs. In addition to the mean, we also draw the 
  min and max of our results. {\it The overlay is robust to attacks
    and churn.}}
\end{figure}

\textbf{Robustness to attacks and churn:} We investigate the robustness 
of the overlay to a massive departure of peers, which can be due to an attack 
or a high churn rate. 
We now consider how we simulate the attack scenario. 
First, we consider the overlay topology shown in Fig.
\ref{connectivity_event10}, which represents a snapshot of the 
topology at time $t=10$. 
Then, we force the most connected peers to leave. For example, assume that 
we want to evaluate the robustness of the overlay to attacks after the departure 
of $30\%$ of the peers. In this case, we identify the $30\%$ most connected peers in 
the overlay and we disconnect them from the overlay. Forcing a peer to leave means that 
we remove all connections between this peer and the rest of the torrent. 
Once these peers are disconnected, we check whether the overlay becomes partitioned, i.e., 
includes more than one partition. By varying the percentage of peers that we force to leave, 
we are able to explore the robustness limits of the BitTorrent overlay. 

To simulate a churn rate, we proceed similarly as for the case of an attack. The only difference 
is that, the peers that we force to leave are selected randomly instead of the most connected ones. 

Fig. \ref{robustness_basic} shows that \bit{}'s overlay is 
robust to attacks and churn. Indeed, the overlay stays connected,
i.e., there is a single partition, when up to $80\%$ of the peers
leave due to an attack or to churn rate. 
When more than $85\%$ of the peers leave the torrent, partitions appear. However, 
there is one major partition that includes most of the peers and a few others 
with one peer each.  

For example, when $95\%$ of the peers leave the torrent due to an attack, the result 
of $1$ run produced $18$ partitions. More precisely, we had $1$ partition that included 
$23$ peers, $2$ partitions that included each $7$ peers, $2$ other partitions with $2$ peers 
each, and $13$ partitions with $1$ single peer each. 
Similarly, when $95\%$ of the peers leave the torrent due to a high churn rate, the overlay 
was split into $4$ partitions, $1$ partition with $43$ peers, $1$ partition with $5$  peers, 
and $2$ partitions with $1$ peer each. 

In summary, we have seen that the
\averageps{} size is significantly lower than the \maxps{} size, and the peer set
size of a peer depends on its arriving time in the torrent.  In
addition, \bit{} generates an overlay with a short diameter, but this
overlay is not a random graph. However, the overlay is robust to
attacks and churn.

\subsection{Validation with Experiments}
\label{val_exp}

\begin{figure}[tb]
   \begin{minipage}{0.49\textwidth}
      \begin{center}
         \subfigure[Experimental results]{
         \includegraphics[height=0.6\textwidth]{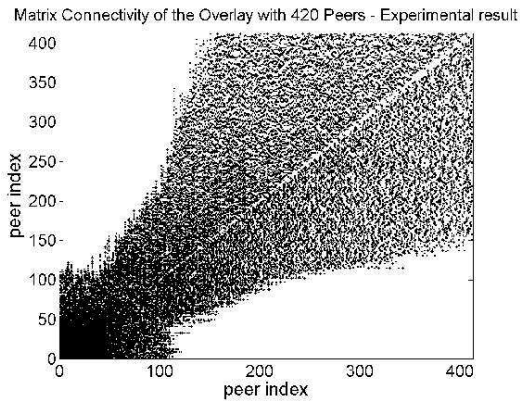}
         \label{experiments_results}}
      \end{center}
  \end{minipage}
   \begin{minipage}{0.49\textwidth}
      \begin{center}
         \subfigure[Simulation results]{
         \includegraphics[height=0.6\textwidth]{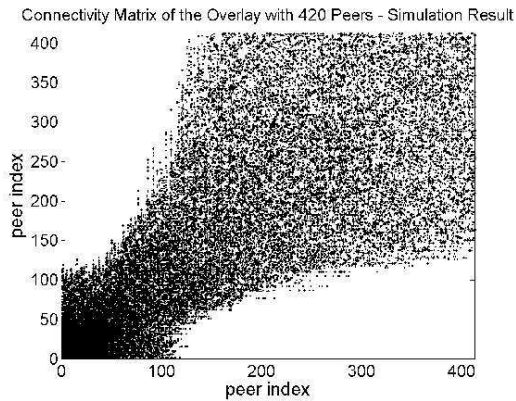}
         \label{simulation_results}}
      \end{center}
  \end{minipage}
\caption{\label{} The connectivity matrix of the \bit{}'s overlay after the arrival of $420$ peers, 
  for a single run as obtained from real experiments and simulations. A dot at (i,j) means that i 
  and j are neighbors.
  {\it Real experiments and simulations show similar properties of the overlay.}}
\end{figure}

To validate our simulation results, we have run real experiments using the mainline client 4.0.2 and its 
tracker implementation that we described in details in Section \ref{simulation_details}. 
In Fig. \ref{experiments_results} and Fig. \ref{simulation_results} we draw the connectivity matrix 
of the overlay as obtained from experiments and simulations respectively. The connectivity matrix is 
computed after the arrival of $420$ peers to the network. As we can see from these two figures, the 
real experiments and the simulations show similar properties of the overlay. 
As a result, our simulator produces accurate results, and as compared to real experiments, it offers much 
more flexibility and allows us to consider larger torrents. Therefore, in the following, we will give 
results only for simulations. 

In the next sections, we will investigate how these results are 
influenced by (1) the size of the torrent, (2) the \maxps{} size, (3) 
the \maxoc{}, and (4) the percentage of \nated{} peers. 
We will also discuss in Section~\ref{sec:impact_pex}
how peer exchange impacts those results.

\subsection{Varying the Number of Peers}

The torrent for the initial scenario is of moderate size. Indeed, it
consists of $1867$ peers and of a maximum of $1200$ simultaneous
peers.  To validate how the results of
Section~\ref{basic_scenario} are impacted by larger torrents, we have
considered two other torrent sizes: $5598$ and
$9329$ peers. For the largest torrent, the maximum number of simultaneous
peers is $6282$.

 \begin{figure}[tb]
 \centering
 \includegraphics[height=0.28\textwidth]{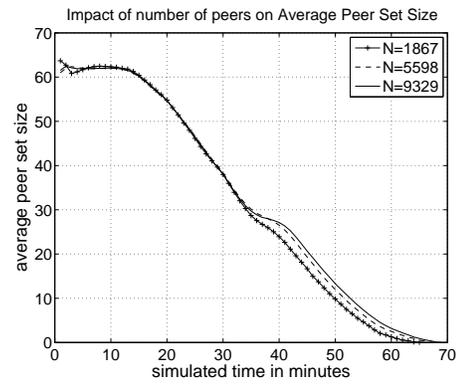}
 \caption{\label{nbpeers_outdegree} The \averageps{} size over time for 
 different torrent sizes, averaged over ten independent runs. 
 {\it The \averageps{} size is independent of the torrent size.}}
 \end{figure}

 \textbf{Average peer set size:} The average peer set size does not depends
 on the torrent size. Indeed, Fig. \ref{nbpeers_outdegree} shows that
 the \averageps{} size is roughly the same for the three torrent sizes.  For
 example, after 10 minutes, when the number of simultaneous peers is
 $1200$ for the smallest torrent and $6282$ for the largest torrent,
 the \averageps{} size is the same for the three torrents.  The evolution of
 the peer set size for the three torrents is similar to the once
 presented in Fig.  \ref{outdegree_distribution}. The \peerset{} size
 decreases from $80$ for the peers that join the torrent early to
 around $40$ for the peers that arrive toward the end.

 \textbf{Convergence speed:} According to section \ref{discussion_convergence}, 
 the convergence speed of the peer set decreases when the torrent size increases 
 as shown in Fig. \ref{fig:convergence_time}. 

 \begin{figure}[tb]
 \centering
 \includegraphics[height=0.28\textwidth]{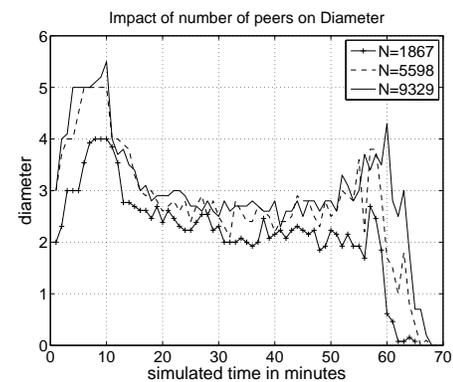}
 \caption{\label{nbpeers_diameter} The diameter over time for different
   torrent sizes, averaged over ten independent runs. 
   {\it The diameter increases slowly with the torrent size.}}
 \end{figure}

 \textbf{Diameter of the overlay:} The diameter of the overlay
 increases slowly with the torrent size. As we can observe in Fig.
 \ref{nbpeers_diameter}, after 10 minutes, the diameter of the overlay is
 $4$ for the smallest torrent and $5.5$ for the largest torrent.  The
 connectivity matrix presents the same characteristics as in
 Fig.~\ref{connectivity_event10} for the three torrents.

 \begin{figure}[tb]
 \centering
 \includegraphics[height=0.28\textwidth]{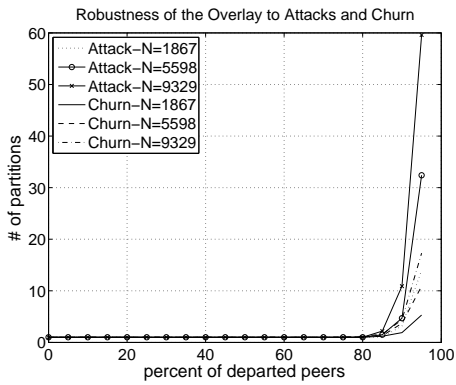}
\caption{\label{impact_attack_churn_N} The robustness of the overlay under churn and attacks, 
averaged over ten different experiments. 
\it{The overlay shows similar robustness for the different torrent sizes.}}
 \end{figure}

 \textbf{Robustness to attacks and churn:} Fig. \ref{impact_attack_churn_N} shows that the 
 robustness of the overlay is independent of the torrent size. In particular, the 
 three overlays stay connected for up to $85\%$ of peer departures. Then, the overlay is
 partitioned with a single large partition and several partitions with a few peers. 
 To show the similarity at the robustness level for the different torrent sizes, we now 
 analyze the number of partitions that we obtain with the attack scenario after the 
 departure of $90\%$ of the most connected peers. 
 Our results show that the torrent of $1867$ peers becomes partitioned into $4$ partitions 
 with one partition of $96$ peers and three others of one single peer each. 
 Similarly, the torrent of $5598$ peers becomes partitioned into $5$ partitions with one 
 partition of $294$ peers and four others of one single peer each. Finally, the torrent of 
 $9329$ peers becomes partitioned into $9$ partitions with one partition of $498$ peers, 
 another partition of two peers, and seven other partitions of one peer each. 
 We found the same tendency for the churn scenario. 

 In summary, we have investigated the impact of the torrent size on the
 properties of the overlay formed by \bit{}. We have found that the
 results obtained for the initial scenario still hold for larger
 torrents. Therefore, and in order to reduce significantly the run time 
 of these simulations, in the following, we will focus on a torrent with 
 $1867$ peers. 

\subsection{Impact of the Maximum Peer Set Size}
\label{impact_maxps}

The maximum peer set size is usually set to $80$. However, some clients choose
higher values of this parameter, e.g., mainline 5.x \cite{bit_site} has a \maxps{} 
size set to $200$. In this section, we evaluate the impact of this parameter 
on the properties of the overlay.

We run simulations with a \maxps{} size $\Delta$ varying from $20$ to $200$.
For each value of $\Delta$, we set the \maxoc{} $O_{max}$ to
$\frac{\Delta}{2}$, the \returned{} to $\frac{\Delta + O_{max}}{2}$,
and the minimum number of neighbors $\delta$ to $20$. Then, we evaluate
the overlay after $10$ simulated minutes, because it is the time at which the
number of simultaneous peers in the torrent reaches its upper bound of $1200$ peers.

Note that there is no specific rule to set the value of the \returned{} $\sigma$ 
when we change the \maxps{} size $\Delta$ and \maxoc{} $O_{max}$. 
Intuitively, $\sigma$ should be larger than $O_{max}$. The reason is that each peer $P_i$ 
seeks to initiate $O_{max}$ connections, which is only possible if the tracker provides 
$P_i$ with the addresses of at least $O_{max}$ other peers in the torrent. Yet, the value of 
$\sigma$ should not be much larger than $O_{max}$ in order not to increase the load on the 
tracker. We tried several values of $\sigma$ between $O_{max}$ and $\Delta$, but we obtained 
similar results. Therefore, we decided to set $\sigma$ to $\frac{\Delta + O_{max}}{2}$. 
\begin{figure}[tb]
\centering
\includegraphics[height=0.28\textwidth]{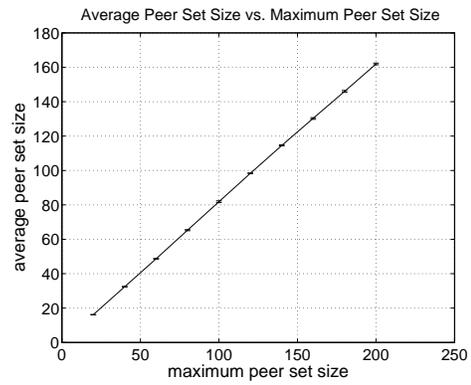}
\caption{\label{bit_maxps_outdegree} The \averageps{} size as a function of
  the \maxps{} size, averaged over ten independent runs. 
  The error bars indicate the minimum and maximum. 
  {\it The average peer set size is a linear function of the maximum peer set size.}}
\end{figure}

\textbf{Average peer set size:} The \averageps{} size increases linearly with
the \maxps{} size $\Delta$. Indeed, Fig.  \ref{bit_maxps_outdegree} shows
that the \averageps{} size is roughly equal to $\frac{2\cdot\Delta}{3}$. For
instance, for a \maxps{} size of $100$, the \averageps{} size is $65$. 

We found this linear trend in all our simulations, but the slope depends on the
instant at which we perform the measurements. 

\textbf{Convergence speed:} We extend the analysis in
Section~\ref{discussion_convergence} by considering a variable
\maxps{} size  $\Delta$ and a \maxoc{} $O_{max}$.
We rewrite Eq. (\ref{eq:convergence_time}) as follows:
\begin{eqnarray}
\label{eq:convergence_time_general} 
\Delta - O_{max} &=& \sum_{n = N_s+1}^{N_s+K} \frac{O_{max}}{n} \notag \\
\frac{\Delta}{O_{max}} &=& 1 + \sum_{n = N_s+1}^{N_s+K} \frac{1}{n}
\end{eqnarray}
where $\Delta - O_{max}$ is the number of missing connections, i.e.,
the number of incoming connections the peer is still waiting for,
assuming that the peer has succeeded to initiate $O_{max}$ outgoing
connections. $\frac{O_{max}}{n}$ is the probability that a peer
receives a new incoming connection from a new peer given the number
of peers $n$ in the torrent. Recall that $N_s$ represents the
torrent size at the moment of the arrival of peer $P_i$,
and $K$ is the average number of peers that should
arrive after peer $P_i$ in order for this peer to complete its
$\Delta - O_{max}$ missing connections. We assume that no peer leaves the system.
When $O_{max}=\frac{\Delta}{2}$ Eq.~(\ref{eq:convergence_time_general}) is equivalent 
to $1 = \sum_{n = N_s+1}^{N_s+K} \frac{1}{n}$, which is Eq.~(\ref{eq:convergence_time}). 
Therefore, when $O_{max}=\frac{\Delta}{2}$, the convergence speed is independent of
the value of $\Delta$.

 \begin{figure}[tb]
 \centering
 \includegraphics[height=0.28\textwidth]{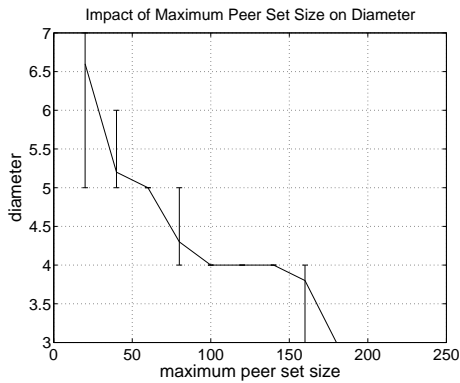}
\caption{\label{bit_maxps_diameter} The diameter of the overlay as a function of
  the \maxps{} size, averaged over ten independent runs. 
  The error bars indicate the minimum and maximum. 
{\it The diameter decreases slowly after a peer set size of $100$.}}
\end{figure}

\textbf{Diameter of the overlay:} The diameter of the overlay decreases slowly 
with the \maxps{} size $\Delta$ as shown in Fig. \ref{bit_maxps_diameter}. 
The diameter is $6.5$ when 
$\Delta$ is $20$, $5.5$ when $\Delta$ is $40$, $4.5$ when $\Delta$ is $80$, 
and $3$ when $\Delta$ is $200$.
However, in contrast to the \averageps{} size, there is no clear trend that can 
be used to predict the diameter as a function of the \maxps{} size.

 \begin{figure}[tb]
 \centering
 \includegraphics[height=0.28\textwidth]{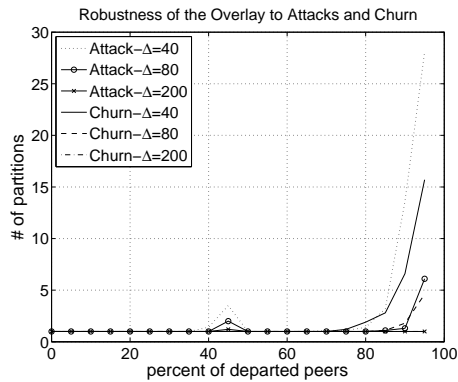}
\caption{\label{impact_attack_churn_maxps} The robustness of the overlay under churn and attacks, 
averaged over ten different experiments. 
\it{The robustness of the overlay increases with the \maxps{} size.}}
 \end{figure}

\textbf{Robustness to attacks and churn:} The robustness of the overlay
increases with the \maxps{} size $\Delta$ as shown in Fig \ref{impact_attack_churn_maxps}. 
For example, when we set $\Delta$ to $40$, $80$, and $200$, the overlay is not partitioned 
for up to respectively $65\%$, $80\%$, and $95\%$ of peers that leave. There is
no discernible distinction between leaves due to churn or attacks. 

\begin{figure}[!tb]
 \centering
 \includegraphics[height=0.28\textwidth]{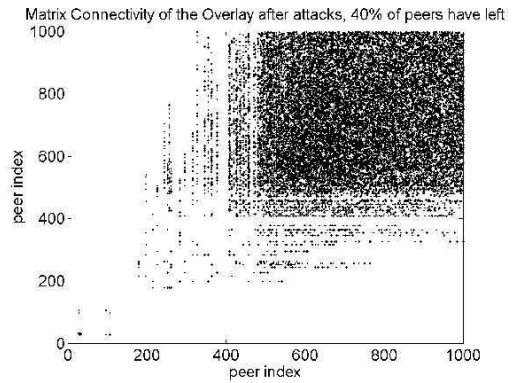}
\caption{\label{basic_overlay_attack45} The connectivity matrix of the overlay after forcing $45\%$ 
of the most connected peers to leave the torrent, for a single run. A dot at (i,j) means that i and 
j are neighbors.
\it{The first peers are disconnected from the rest of torrent.}}
 \end{figure}

However, if we carefully look at Fig. \ref{impact_attack_churn_maxps}, we can see that the 
attack scenario produces partitions when the percentage of departed peers is at $45\%$. To understand 
this behavior, we plot in Fig. \ref{basic_overlay_attack45} the connectivity matrix of the overlay 
after attacking $45\%$ of the peers. Fig. \ref{basic_overlay_attack45} shows that there are around 
$12$ peers that are disconnected from the rest of the torrent. 

From our previous results, we know that the peers that arrive at the beginning of the torrent are the 
most connected ones and highly connected among each other (see Fig. \ref{connectivity_event10}). In 
addition, these first peers are the most concerned ones by the attack. Recall that the attack scenario 
forces the most connected peers to leave the torrent. 
Thus, with a $45\%$ of departed peers, only very few of those first peers will remain present in the torrent 
after the attack. These few peers will have a very few neighbors and will become disconnected from other peers. 
However, if we consider a percentage of departed peers of $50\%$ instead of $45\%$, more peers will leave the 
torrent and those $12$ peers will disappear. This means that the torrent will be connected again. 
In contrast, if  we consider a percentage of departed peers of $40\%$ instead of $45\%$, there will be more peers 
present in the torrent after the attack, which helps the torrent to remain connected.

In summary, the \maxps{} size does not have a major impact on the
properties of the overlay, as long as the \maxps{} size is large
enough to have a small diameter. In our simulations, we do not see a
major difference in the overlay properties between a \maxps{} size of
$80$ and $200$. However, as the \maxps{} size increases linearly the
average peer set size, it also increases the speed of replication of
the pieces (according to
Section~\ref{discussion_outdegree}). Therefore, the main reason to
increase the \maxps{} size is to improve the speed of
replication. But, there is a tradeoff, as a larger \maxps{} size
increases the load on each client due to the larger number of TCP
connections to maintain and due to the signaling overhead per
connection.

\subsection{Impact of the Maximum Number of Outgoing Connections}
\label{impact_maxoc}
The \maxoc{} $O_{max}$ is critical to the properties of the
overlay. Indeed, when $O_{max}$ is close to the \maxps{} size $\Delta$, the
peer set size will converge fast to $\Delta$, but new peers will find
few peers with available incoming connections, hence a larger
diameter. 

In this section, we evaluate the impact of $O_{max}$ on the overlay
properties. For the simulations, we set $\Delta$ to $80$, the minimum
number of neighbors to $20$, and we vary $O_{max}$ from $5$ to $80$
with a step of $5$. For each value of $\Delta$ and $O_{max}$, the
number of peers returned by the tracker is equal to $\frac{\Delta +
  O_{max}}{2}$.

\begin{figure}[tb]
\centering
\includegraphics[height=0.28\textwidth]{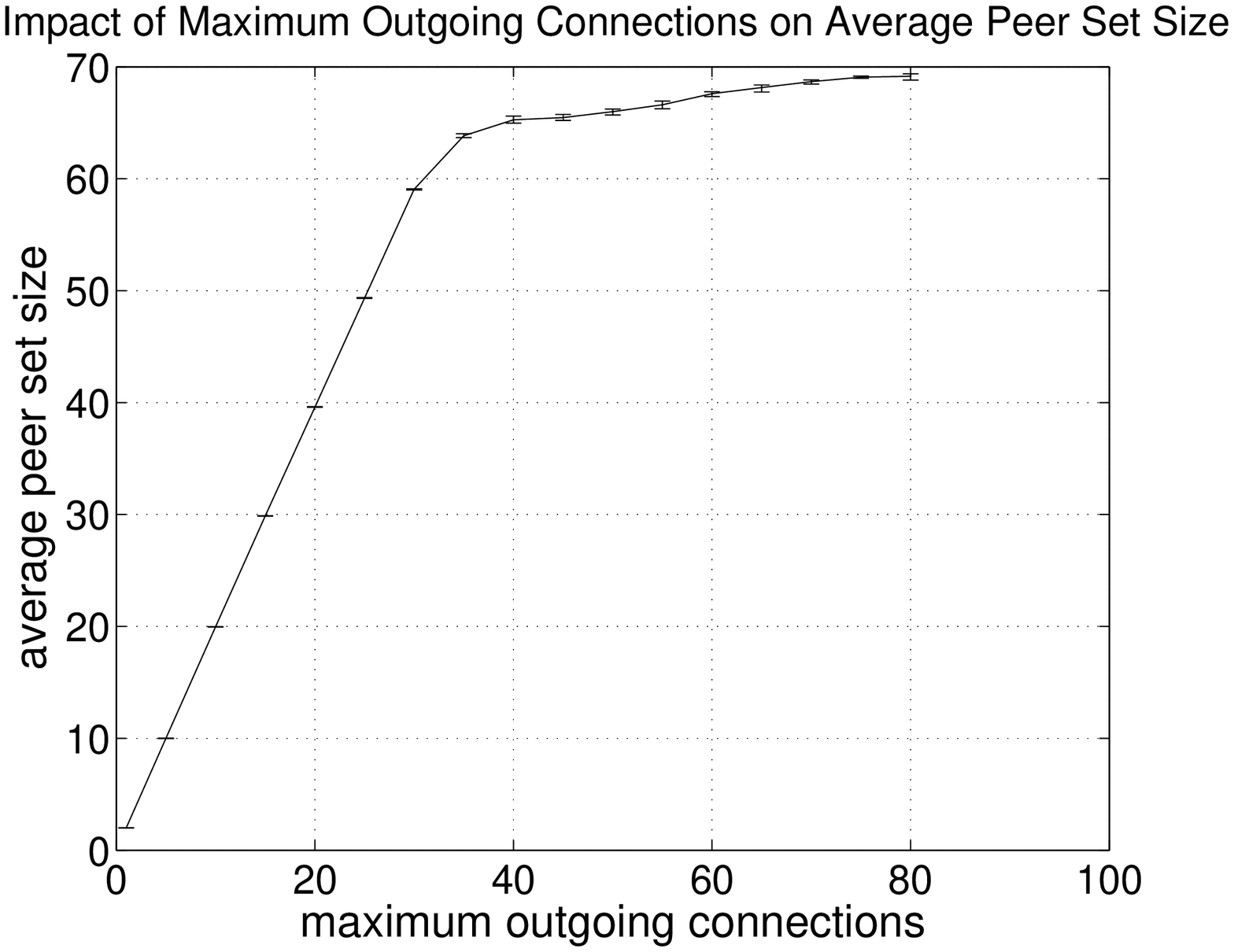}
\caption{\label{tracker_outdegree_ratio_maxps80} The \averageps{} size
  as a function of the \maxoc{}, averaged over ten independent runs. 
  The error bars indicate the minimum and maximum. 
  {\it The \averageps{} size increases slowly when the number of \maxoc{} is larger than 
    $\frac{\Delta}{2}$.}}
\end{figure}

\textbf{Average peer set size:}
Fig. \ref{tracker_outdegree_ratio_maxps80} shows the evolution of the
\averageps{} size as a function of the \maxoc{}. We see that the
\averageps{} size increases fast with $O_{max}$ when $O_{max}$ is smaller
than $\frac{\Delta}{2}$, and it increases slowly with $O_{max}$ when
$O_{max}$ is larger than $\frac{\Delta}{2}$. We notice that, a small $O_{max}$
leads to a small \averageps{} size. For example, when $O_{max}$ is
equal to $5$, the \averageps{} size is around $10$.

\begin{figure}[tb]
\centering
\includegraphics[height=0.28\textwidth]{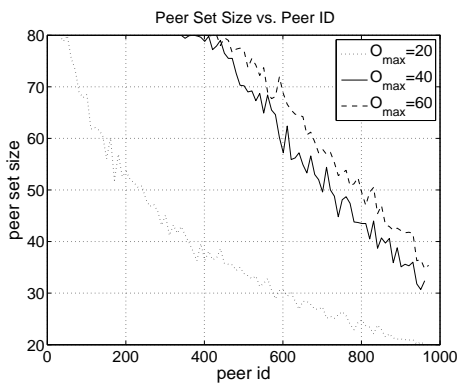}
\caption{\label{outdegree_dist_maxoc} The peer set size vs. the peer id, at time 
 $t=10$ minutes, averaged over ten independent simulations. 
  {\it The peer set size increases slowly when the number of \maxoc{} is larger than 
    $\frac{\Delta}{2}$.}}
\end{figure}

\textbf{Convergence speed:} In Fig. \ref{outdegree_dist_maxoc} we plot the peer set 
size of each peer at time $t=10$ minutes as a function of the peer id for three values 
of the \maxoc{}. These outdegree distributions of peers reflect the convergence speed of 
peers towards their \maxps{} size. 
As we can see from Fig. \ref{outdegree_dist_maxoc}, 
the peer set size of the different peers improves with $O_{max}$ when $O_{max}$ is 
smaller than $\frac{\Delta}{2}$, and it increases slowly with $O_{max}$ when $O_{max}$ 
is larger than $\frac{\Delta}{2}$. 
For example, after the arrival of the first $1000$ peers $P_1, \dots, P_{1000}$, 
peer $P_{500}$ has a peer set size of $34$ 
when $O_{max}$ is at $20$ and a peer set size of $70$ when $O_{max}$ is at $40$. However, 
when we increase $O_{max}$ from $40$ to $60$, the peer set size of $P_{500}$ increases from 
$70$ to $76$ only. This result means that, when we increase $O_{max}$ beyond $\frac{\Delta}{2}$, 
the convergence speed of a peer towards its \maxps{} increases slowly. 
We explain this conclusion as follows. According to Eq.~(\ref{eq:convergence_time_general}), 
when $O_{max}$ increases, $K$ decreases, where $K$ is the number of peers that should arrive 
after a peer $P_i$, so that $P_i$ reaches its maximum peer set size. Indeed,
when $O_{max}$ increases, the probability that a peer receives
incoming connections from new peers increases too.  However, to derive
Eq.~(\ref{eq:convergence_time_general}), we assumed that a peer
succeeds to establish all its allowed $O_{max}$ outgoing connections
and that the number of connections it misses is $\Delta -
O_{max}$. This is the most optimistic case, and it is not true when
$O_{max}$ is larger than $\frac{\Delta}{2}$.
Indeed, in that case, peers that
arrive at the beginning of the torrent are able to establish a lot of
connections among themselves and reach fast their \maxps{} size. However,
those peers leave few rooms for incoming connections, as $O_{max}$ is
close to the \maxps{} size $\Delta$. Therefore, peers that join later the
torrent will not be able to establish $O_{max}$ outgoing connections,
which results in a larger number than $\Delta - O_{max}$ of missing connections.
As a consequence, the increase in the probability that a peer is
selected by new arriving peers is compensated by the increase in the
number of missing connections. 

 \begin{figure}[tb]
   \begin{minipage}{0.49\textwidth}
      \begin{center}
         \subfigure[\maxoc{} of $70$]{
         \includegraphics[height=0.6\textwidth]{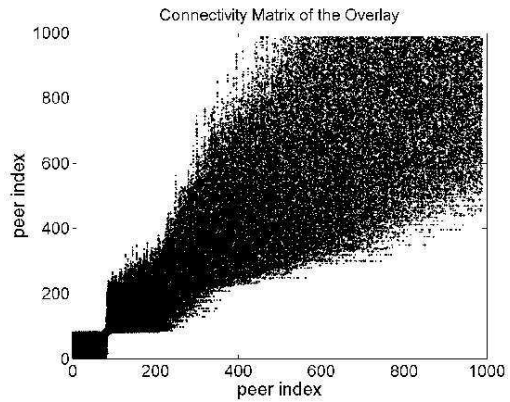}
         \label{tracker_N1867_event10_maxoc70_maxps80_connectivity}}
      \end{center}
  \end{minipage}
   \begin{minipage}{0.49\textwidth}
      \begin{center}
         \subfigure[\maxoc{} of $80$]{
         \includegraphics[height=0.6\textwidth]{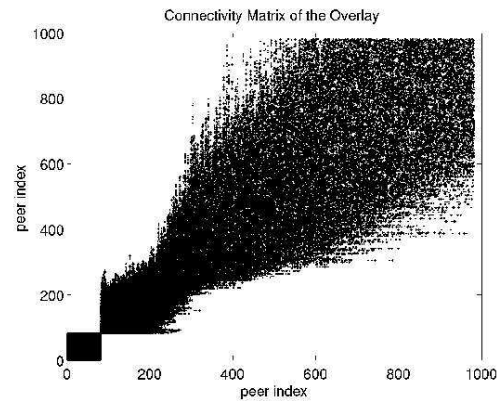}
         \label{tracker_N1867_event10_maxoc80_maxps80_connectivity}}
      \end{center}
  \end{minipage}
\caption{\label{tracker_N1867_event10_maxoc70_maxps80_connectivity_all}
  The connectivity matrix of the \bit{} overlay after 10 minutes for a 
  \maxoc{} of $70$ and $80$, for a single run. A dot at (i,j) means that i
  and j are neighbors. {\it When we increase $O_{max}$ beyond $\frac{\Delta}{2}$, 
  the connectivity matrix becomes more narrow around peer index 80.}}
\end{figure}

\begin{figure}[tb]
\centering
\includegraphics[height=0.28\textwidth]{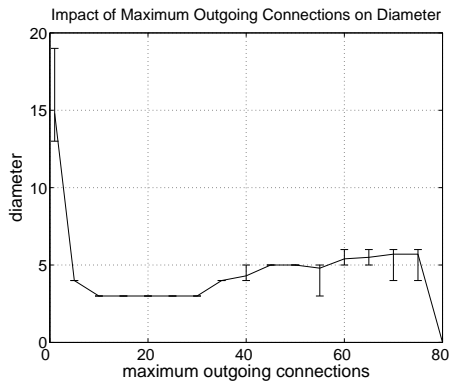}
\caption{\label{impact_maxoc_diameter} The diameter of the overlay as a function 
  of the \maxoc{}, averaged over ten independent runs. 
  The error bars indicate the minimum and maximum. 
  {\it Taking a \maxoc{} $O_{max}$ larger than $\frac{\Delta}{2}$ increases the 
  diameter of the overlay.}}
\end{figure}

\textbf{Diameter of the overlay:} Taking a \maxoc{} $O_{max}$ larger than
$\frac{\Delta}{2}$ increases the diameter of the overlay. As we can see in Fig. 
\ref{impact_maxoc_diameter}, for $O_{max}$ equal to $40$ (respectively $70$), the diameter of
the overlay is equal to $4$ (respectively $5.5$). When $O_{max}$ is equal to
$\Delta$, i.e., 80,  the overlay is partitioned. 
If we focus on the connectivity matrix of the overlay, we observe how
the overlay gets partitioned into two partitions. Indeed,
Fig.~\ref{tracker_N1867_event10_maxoc70_maxps80_connectivity} shows
that when $O_{max}$ is equal to 70, the connectivity matrix becomes narrow
around peer index 80. This results in the first $80$ peers in the torrent
being highly connected among themselves with $3115$ connections, and poorly 
connected with the rest of the torrent with $170$ connections. 
When $O_{max}$ is equal to 80, the first 80 peers
become disconnected from the rest of the torrent. This might be a
major issue if the source of the torrent is among those 80 peers,
which is the regular case. 

 \begin{figure}[tb]
 \centering
 \includegraphics[height=0.28\textwidth]{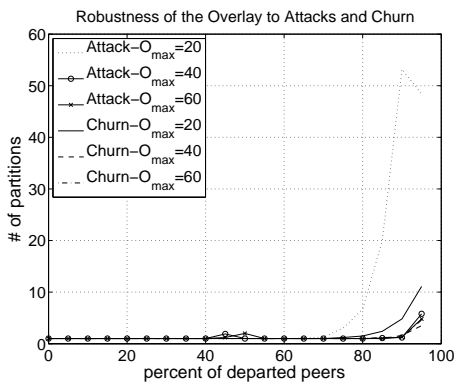}
\caption{\label{impact_attack_churn_maxoc} The robustness of the overlay under churn and attacks, 
averaged over ten different experiments. 
\it{The robustness of the overlay increases slightly with the \maxps{} size.}}
 \end{figure}

\textbf{Robustness to attacks and churn:} Fig. \ref{impact_attack_churn_maxoc} draws the 
robustness of the
overlay with a \maxoc{} $O_{max}$ set to $20$, $40$, and $60$. We observe that 
large values of $O_{max}$ make the overlay slightly more robust to attacks and 
churn. For example, in case of an attack, when setting $O_{max}$ to $20$, the partitions appear 
after the departure of $60\%$ of the peers. In contrast, when setting 
$O_{max}$ to $40$ or $60$, the partitions appear after the departure 
of $85\%$ of the peers. 

In Fig. \ref{impact_attack_churn_maxoc}, we can also show that, when we set $O_{max}$ of $20$, the number 
of partitions decreases at the end of the curve. 
The reason is that, when we force $90\%$ or more of the most connected peers to leave the network, 
a $O_{max}$ of $20$ will produce a lot of partitions with a very few peers each. Thus, increasing 
the number of departing peers removes those ``one single peer partitions''. 

Note that, when $O_{max}$ is set to $40$ (respectively to $60$), the attack scenario produces partitions when the 
percentage of departed peers is at $45\%$ (respectively $50\%$). The reason is that, among the most connected peers, 
the remaining peers are very few to stay connected with the rest of the torrent. 

In summary, setting the \maxoc{} $O_{max}$ to $\frac{\Delta}{2}$ is a
good tradeoff between the average peer set size and the diameter of the
overlay.

\subsection{Impact of the Number of \nated{} Peers}
\label{sec:impact-number-nated}
In this section, we discuss the impact of the percentage of \nated{}
peers on the overlay properties. When a peer is behind a NAT, it
cannot receive incoming connections from other peers in the
torrent. However, it can initiate outgoing connections to non \nated{}
ones. 
For the simulations, we set the \maxps{} size to $80$, the
\maxoc{} to $40$, the minimum number of neighbors $\delta$ to $20$,
and the number of returned peers by the tracker to $50$. Then, we vary
the percentage of \nated{} peers from $0\%$ to $90\%$ with steps of $10$.

 \begin{figure}[tb]
 \centering
 \includegraphics[height=0.28\textwidth]{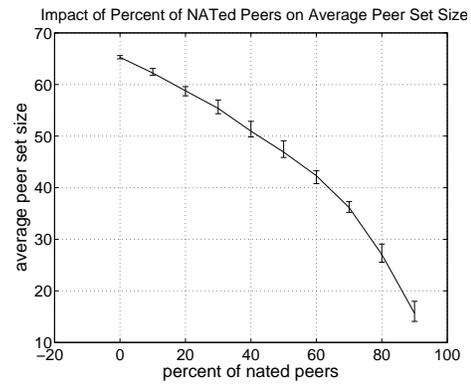}
 \caption{\label{outdegree_nated} The \averageps{} size vs. the percentage of \nated{} 
 peers, averaged over $10$ independent runs. 
  The error bars indicate the minimum and maximum. 
 \it{The \averageps{} decreases linearly with the percentage of \nated{} peers.}}
 \end{figure}

\textbf{Average peer set size:} \bit{} mitigates very efficiently the
impact of the \nated{} peers on the overlay. For example, we see in 
Fig. \ref{outdegree_nated} that as we increase 
the percentage of \nated{} peers from $0$ to $30\%$, the \averageps{} 
size is reduced from $65$ to $55$. Indeed, the \averageps{}
size decreases slowly with the percentage of \nated{} peers. However, the 
slope of the curve becomes much sharper when the percentage of \nated{} peers 
exceeds $60\%$.

\textbf{Convergence speed:} The convergence speed that we derive in
Eq. (\ref{eq:convergence_time}) holds for non \nated{} peers. When a
peer is \nated{}, it will establish at most $O_{max}$ outgoing
connections, which is the higher bound for its \maxps{} size. 

 \begin{figure}[tb]
 \centering
 \includegraphics[height=0.28\textwidth]{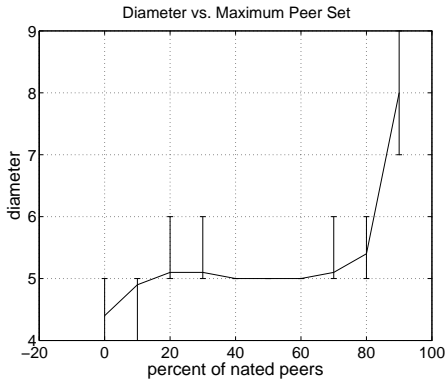}
 \caption{\label{diameter_nated} The diameter of the overlay vs. the percentage of 
 \nated{} peers,  averaged over $10$ independent runs.
  The error bars indicate the minimum and maximum. 
 \it{The diameter increases slowly with the number of \nated{} peers.}}
 \end{figure}

\textbf{Diameter of the overlay:} \nated{} peers do not make the diameter significantly larger. 
For example, as shown in Fig. \ref{diameter_nated}, when 
$10\%$ (respectively $80\%$ of the peers are \nated{}, 
the average value of the diameter is at $4.5$ (respectively $5.5$). 
However, in extreme cases where $90\%$ of the peers are \nated{}, the diameter can reach $9$. 

Note that, \nated{} peers may cause partitions. For example, assume that peer 
$P_i$ is \nated{}. It may happen that, out of the peers returned by the tracker to $P_i$, no one 
has room for more connections. As a result, $P_i$ will not be able to establish any outgoing connections. 
In addition, and because it is \nated{}, $P_i$ cannot receive connections from other peers. As a result, peer 
$P_i$ will be isolated alone (disconnected from the rest of the torrent) until it contacts the tracker 
again and discovers more peers. This behavior becomes more common as the percentage of \nated{} peers increases. 

 \begin{figure}[tb]
 \centering
 \includegraphics[height=0.28\textwidth]{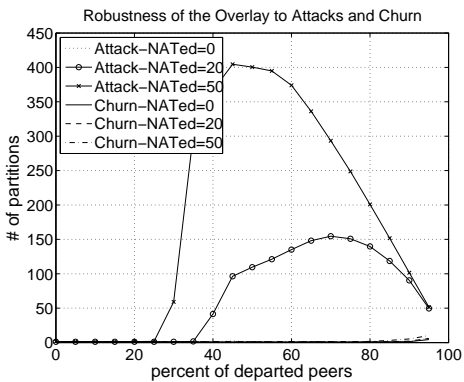}
\caption{\label{impact_attack_churn_nated} Robustness of the overlay with
  different percentage of \nated{} peers. {\it The robustness to churn
  does not depend on the percentage of \nated{} peers, but the overlay
is not robust to attacks when there are \nated{} peers.}}
\end{figure}

\textbf{Robustness to attacks and churn:} 
The robustness of the overlay to high churn rates does not depend on the presence 
of \nated{} peers. Indeed, Fig.~\ref{impact_attack_churn_nated} shows that the overlay
stays connected when up to $80\%$ of peers leave the torrent. However,
the overlay is not robust to attacks when there is a large number of
\nated{} peers. Indeed, Fig.~\ref{impact_attack_churn_nated} shows that when
there are $50\%$ of \nated{} peers, the overlay starts to be
partitioned when $25\%$ of the peers leave due to an attack. We see
that the number of partitions decreases for a large percentage of
departing peers, because as there are many small partitions,
increasing the number of departing peers removes those partitions. 

In summary, \nated{} peers decrease significantly the robustness of 
the overlay to attacks. 

\section{Impact of Peer Exchange on the Overlay}
\label{sec:impact_pex}

We have seen that \bit{} generates overlays with a short diameter 
that are robust to churn and attacks. However, the time for a peer to
reach its \maxps{} size depends on the torrent size and peer arrival rate. 
One way to reach faster the \maxps{} size is to increase the number of
requests to the tracker in order to discover more peers and establish 
more connections. However, such requests increase the load on
the tracker, whereas the tracker is known to have scarce resources
\cite{NEGL07_INFOCOM,POUW05_IPTPS}. 

In Mai $2005$, Azureus 2.3.0.0 \cite{azureus} introduced a new feature, namely 
\textit{peer exchange} (\pex{}), 
where neighbors periodically exchange their list of neighbors. For example, 
assume that peers $P_i$ and $P_j$ are neighbors. Then, every minute, $P_i$ sends 
its list of neighbors to $P_j$ and vice versa. As a result, each peer knows its 
neighbors and the neighbors of its neighbors. The intuition behind \pex{} is that 
peers will be able to discover fast a lot of peers and consequently achieve a 
larger peer set size. 

Note that the results that we have given in previous sections are for the case of 
BitTorrent without \pex{}, e.g., the official BitTorrent client \cite{bit_site}. 
In this section, we extend our work and analyze how \pex{} impacts the overlay 
topology of BitTorrent. \pex{} is becoming very popular and, in addition 
to Azureus, it is now implemented in several other P2P clients including KTorrent, 
libtorrent, $\mu$Torrent, or BitComet, but with incompatible implementations. 
To the best of our knowledge, the impact of \pex{} has never been discussed 
previously.

\subsection{Simulating \pex{}}

To evaluate the impact of \pex{} on the overlay topology, we added this feature to 
our simulator exactly as it is implemented in Azureus. 
Concerning the communications between peers and the tracker, 
all what we described in Section \ref{simulation_details} is still valid. 
That is, the tracker keeps two lists, $\natedlist$ and $\nonnatedlist$ peers. And, 
when a peer $P_i$ joins the torrent, it gets from the tracker up to $\sigma$ 
peers randomly selected from $\nonnatedlist$. Then, $P_i$ stores those IP addresses 
in its $L_{tracker}^{P_i}$ list and initiates sequentially up to $O_{max}$ connections 
to those peers. Moreover, $P_i$ will be added at the tracker to either $\natedlist$ or 
$\nonnatedlist$.

We now explain the modifications that we made in our simulator. 
Assume that, at time $t=0$, a connection has been established between 
two peers $P_i$ and $P_j$. Just after, $P_i$ sends its list 
of neighbors to $P_j$ and vice versa. 
Then, every $1$ simulated minute, $P_i$ and $P_j$ repeat this exchange process. 

Assume now that, after performing \pex{} with its neighbor $P_j$, $P_i$ discovers peer $P_k$. 
Then, $P_i$ checks whether (1) it already has a connection with $P_k$ or (2) it already knows 
$P_k$ from the tracker. If none of these two 
conditions holds true, then $P_i$ adds $P_k$ to the list $L_{pex}^{P_i}$ of peers it discovered 
through \pex{}. 
Note that, $P_i$ may receive the IP address of $P_k$ from many neighbors. In this case, $P_k$ 
will appear only once in $L_{pex}^{P_i}$. 

Note that, when establishing connections, peers discovered from the tracker are given more 
priority. For example, 
assume that peer $P_i$ decides to initiate a new connection, which can be due to the departure of 
one of its neighbors or after discovering new peers. In this case, $P_i$ contacts first the peers it 
has discovered from the tracker. If none of those peers accepts the connection request, $P_i$ 
contacts the peers that it discovered through \pex{}. 

\subsection{Analysis of \pex{}}
We implement \pex{} in our simulator and run simulations with the
following parameters. We set the \maxps{} size to $80$, the \maxoc{}
to $40$, the minimum number of neighbors to $20$, and the number of
peers returned by the tracker to $50$. 
However, due to the gossiping messages between peers, the \pex{} feature makes 
our simulator very slow. In order to save time, we run simulations for a 
torrent of $1000$ peers that arrive to the torrent within the first $60$ simulated 
minutes according to Eq. \ref{arrival_eq}. The departure of peers is scheduled during 
the next simulated $60$ minutes, and it follows a random uniform distribution. 
For example, if peer $A$ arrives at time $t=30$, it will leave the network at a random 
time uniformly selected between $t=60$ and $t=120$.
Still, this torrent allows us to understand how the overlay is constructed with \pex{} and 
how it evolves as peers join and leave. 

\begin{figure}[tb]
\centering
\includegraphics[height=0.28\textwidth]{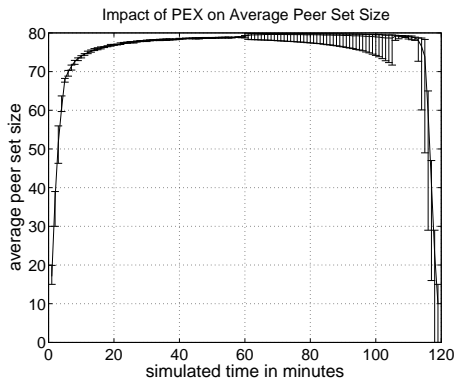}
\caption{\label{pex_outdegree} Evolution of the outdegree of peers with time with \pex{}, 
 averaged over ten independent runs. 
  The error bars indicate the minimum and maximum. 
 {\it \pex{} produces an average peer set size that is very close to the \maxps{} size.}}
\end{figure}

\textbf{Average peer set size:} \pex{} meets its intended goal and
permits peers to be at their \maxps{} size most of the time as shown in 
Fig. \ref{pex_outdegree}. Moreover,
\pex{} prevents the \averageps{} size from decreasing in case of a
massive departure of peers. Indeed, when a peer loses a connection due
to the departure of a neighbor, it can replace it by a new connection
to one of its neighbors' neighbors. As a result, the \averageps{} size
stays at its maximum value of $80$ as long as there are $80$ peers in
the torrent.

\textbf{Convergence speed:} Each peer reaches its maximum peer set
size within a few gossiping period (that we set to one minute in our
simulations). 

\begin{figure}[tb]
\centering
\includegraphics[height=0.28\textwidth]{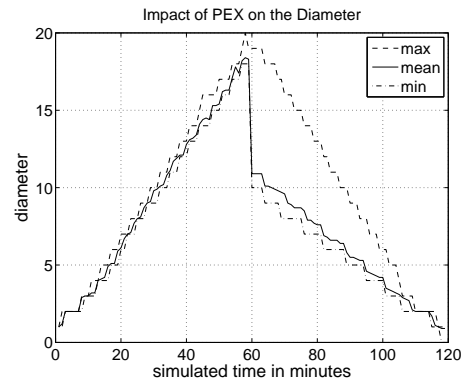}
\caption{\label{pex_diameter} Evolution of the overlay diameter with
  time with \pex{}, averaged over ten independent runs. We plot the mean, max, and min values. 
 {\it \pex{} generates overlays with very large diameter.}}
\end{figure}

\begin{figure}[tb]
\centering
\includegraphics[height=0.28\textwidth, angle=90]{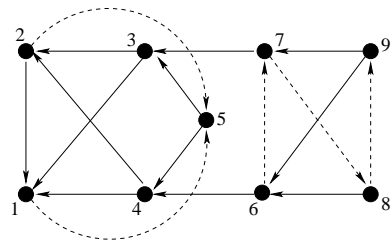}
\caption{\label{pex_schema} The topology evolution with \pex{} with 9
  peers. Connections established with the tracker information are shown
  with solid lines, and connection established with \pex{} are shown with dashed
  lines.}
\end{figure}

\textbf{Diameter of the overlay:} The increase in the \averageps{}
size comes at the expense of a larger diameter. As we see in
Fig.~\ref{pex_diameter}, the maximum value of the diameter reaches $20$ 
when the number of peers in the torrent is $1000$.

To explain why \pex{} produces such a long diameter, we plot in
Fig.~\ref{pex_schema} the evolution of a torrent with 9 peers ($P_1,
\dots, P_9$) that arrive sequentially, one every 1 unit of time.  In
this example, we set the \maxps{} size to 4, the \maxoc{} to 2, and
the \returned{} to 2. At time $t=0$, there is only peer $P_1$. At time
$t=2$, $P_2$ joins the torrent and connects to $P_1$. At time $t=3$,
$P_3$ arrives and connects to $P_1$ and $P_2$. Then, $P_4$ arrives and
connects to two existing peers selected at random, say $P_1$ and
$P_2$. At the end of time $t=3$, \pex{} has not yet been used.
At time $t=4$, $P_5$ arrives and connects to $P_3$ and $P_4$, which in
turn tell $P_1$ and $P_2$ about this new neighbor. At this time, $P_1$
and $P_2$ each has a room for one more outgoing connection and both
connect to $P_5$.  Thus, at time $t=4$, only peers $P_3$ and
$P_4$ can accept new incoming connections, i.e., $P_1$, $P_2$, and
$P_5$ have already reached the maximum number of connections.
At time $t=5$, peer $P_6$ joins the torrent and gets from the tracker
the addresses of $P_1$ and $P_4$. However, $P_6$ can only initiate a
connection to $P_4$, as $P_5$ has already reached its \maxps{} size. 
Peer $P_7$ arrives at time $t=6$, gets the 
addresses of $P_3$ and $P_5$ and initiates one connection to
$P_3$. Then, $P_8$ joins the torrent at time $t=7$, obtains the
addresses of $P_2$ and $P_6$ and initiates only one connection to
$P_6$. At time $t=8$, peer $P_9$ arrives and gets the addresses of
$P_6$ and $P_7$. Given that these two peers have not reached their
\maxps{} size,
$P_9$ succeeds to connect to both of them.  Afterward, $P_6$ tells its
neighbor $P_8$ about $P_9$, thus a new connection is initiated from
$P_8$ to $P_9$. Then, $P_9$ tells $P_7$ about $P_8$ and a new
connection is initiated from $P_7$ to $P_8$.

As we can see, \pex{} tries to maximize the number of outgoing
connections at each peer. Peers keep on gossiping and whenever they
discover new peers, they establish new connections if they still have
room for. However, the disadvantage is that peers that arrive at the
beginning establish a lot of connections among each other and leave
only a few free connections for the peers that arrive afterward. In
the example that we consider here, $P_1, \dots, P_5$ are highly
interconnected and they leave only two connections for next
peers. Similarly, when peers $P_6, \dots, P_9$ arrive, they connect to
the overlay at peers $P_3$ and $P_4$ and then interconnect strongly
with each other. 

\begin{figure}[!t]
\centering
\includegraphics[height=0.28\textwidth]{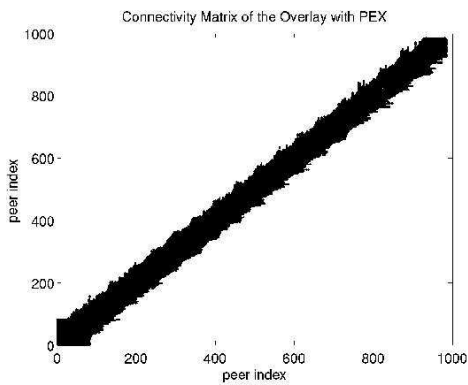}
\caption{\label{pex_connectivity} The connectivity matrix of the overlay distribution with \pex{} 
 after $59$ minutes, for a single run. A dot at (i,j) means that i and j are neighbors. At $t=59$, 
 almost all the $1000$ peers have joined the network and no peer has left yet. 
{\it We observe a chain-like overlay.}}
\end{figure}

As a result, \pex{} leads to a clustering phenomena, where each
cluster contains approximately a number of peers close to the \maxps{}
size. Each cluster exhibits a high intra-cluster connectivity and a
poor inter-cluster connectivity with the cluster that arrives just
before and to the one that arrives just after. To confirm our
analysis, we draw in Fig.~\ref{pex_connectivity} a snapshot of the
connectivity matrix of the overlay distribution after $60$ minutes when
the first $1000$ peers have arrived in the torrent.  The clustering
phenomena appears clearly in the figure, which explains the large diameter 
of the overlay. As we explained in Section~\ref{discussion_diameter}, 
such a chain-like overlay constraints the distribution time in the system to 
be a linear function of the number of clusters.
As compared to the overlay generate by the tracker only, this chain-like 
overlay becomes less efficient when the number of clusters becomes larger than the number 
of pieces. Typical files distributed using \bit{} includes an average of $1000$ pieces. 
In this case, this chain-like overlay will become inefficient when the number of clusters 
is larger than $1000$, i.e., the number of peers is larger than $800.000$ peers. 
Current torrents are much smaller and they rarely exceed 100.000 peers and 
therefore, no one has noticed yet the negative impact of the \pex{} on the download 
time of files. 

Let us now go back to Fig. \ref{pex_diameter}. If we carefully look at this figure, at time 
$t=60$ minutes, the average value of the diameter drops from $18$ to $11$. Actually, at 
$t=60$ minutes, peers start leaving the torrent. At the same time, the last peers to join 
the torrent also arrive at $t=60$ minutes. 
Those departed peers will allow the arriving ones to connect at different levels of the chain 
and not only at the tail, which consequently reduces the diameter of the overlay. 

To better explain this behavior, consider the overlay shown in Fig. \ref{pex_connectivity}. 
In this chain-like overlay, all peers are at their \maxps{} size except those that are at 
the tail. 
Assume now that peers $P_1, \dots, P_{15},$ leave the torrent. Assume that, at the same 
time, $P_{985}, \dots, P_{1000},$ join the torrent. 
In this case, the departed peers $P_1, \dots, P_{15},$ will leave rooms for incoming 
connections inside the first cluster, i.e., at the head of the chain. Thus, the arriving peers 
$P_{985}, \dots, P_{1000},$ will be able to establish connections to the tail as well as 
the head of the chain. As a result, the head and tail of the chain become connected and 
the diameter of the overlay drops by half. Still, the diameter remains very high when compared 
to an overlay generated only by the tracker.

 \begin{figure}[tb]
   \begin{minipage}{0.49\textwidth}
      \begin{center}
         \subfigure[Impact of attack on the \pex{} overlay]{
         \includegraphics[height=0.6\textwidth]{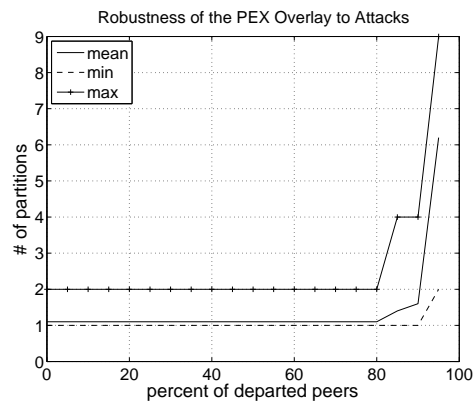}
         \label{impact_attack_pex}}
      \end{center}
  \end{minipage}
   \begin{minipage}{0.49\textwidth}
      \begin{center}
         \subfigure[Impact of churn on the \pex{} overlay]{
         \includegraphics[height=0.6\textwidth]{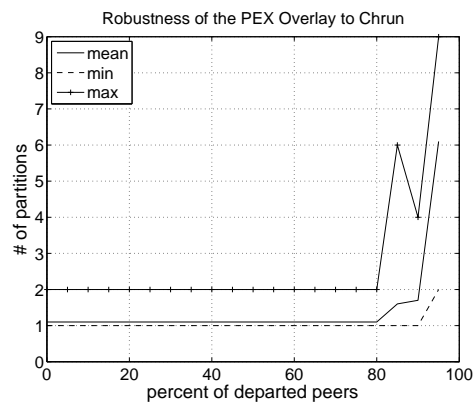}
         \label{impact_churn_pex}}
      \end{center}
  \end{minipage}
\caption{Robustness of the chain-like overlay generated by \pex{}. 
  {\it The \pex{} overlay can be easily partitioned by simply 
attacking a peer, its neighbors, and its neighbors of neighbors.}}
\end{figure}

\textbf{Robustness to attacks and churn:} Surprisingly, the overlay produced with \pex{} 
is robust to churn rate and to the attack that targets the most connect peers. As shown 
in Fig. \ref{impact_attack_pex} and Fig. \ref{impact_churn_pex}, the overlay stays connected 
with up to $80\%$ of the peers leaving the torrent. 

As we can see in Fig. \ref{impact_attack_pex}, up to $80\%$ of departed peers, the attack scenario 
produces a maximum number of partitions of $2$. Actually, out of the ten runs that we performed, we 
obtained partitions with only $1$ run. In particular, we obtained one major partition and a second 
one with only $1$ peer. For example, for $80\%$ of departed rate, we obtained $1$ partition 
that includes $167$ peers and a second partition with only $1$ peer. 
Same conclusions apply on the churn scenario. 

Even though \pex{} shows good robustness to the attack that we have been using 
so far, this chain-like overlay can be easily partitioned by using more sophisticated 
attacks that target a peer, its neighbors, and its neighbors of neighbors. 

In summary, even if \pex{} significantly decreases the time for a peer to reach 
its maximum peer set size, it creates a chain-like overlay that is not robust 
against partitions and whose diameter is large. This large diameter will lead to 
a long download time of files when the number of simultaneous peers is large.
We plan to evaluate how much this overlay impacts the efficiency of the transfer 
when compared to an overlay created only by the tracker.


\section{Discussion}
\label{sec:conc}

\subsection{Summary of our Contributions}
We have conducted a large set of simulations to investigate the properties of 
the overlay formed by \bit{}. Below is a list of our main contributions.

\begin{itemize}

\item First, we have analyzed the relation between the overlay properties and the performance 
of \bit{}. In particular, we have shown that a large peer set size increases the 
efficiency of \bit{}, and that a small diameter is a necessary, but not sufficient, 
condition for this efficiency. 

\item Second, we have shown for the first time that the overlay generated by \bit{} 
is not a random graph, as it is commonly believed. The connectivity of a peer 
with neighbors in the torrent is highly biased by its arriving order in the 
torrent. 
Whereas it is beyond the scope of this study to evaluate the robustness of 
the overlay structure to elaborated attacks, i.e., attacks that do not
only focus on the most connected peers, it is an interesting area for
future research. In particular, it is critical to understand such
issues when a public service is to be built on top of \bit{}.

\item Third, we have evaluated the impact of the maximum peer set size and
of the maximum number of outgoing connections. Whereas there are
several magic numbers in \bit{}, we have identified that the
\maxps{} size is a tradeoff between efficiency and peers overhead, and
we have explained why the maximum number of outgoing connections must
be set to half of the \maxps{} size. 

\item Finally, we have identified two potentially significant problems in
the overlay, which deserve further investigations. We have shown that
a large number of \nated{} peers decrease significantly the robustness
of the overlay to attacks, and we have shown that peer exchange
creates a chain-like overlay that might adversely impact the
efficiency of \bit{}.

\end{itemize}

In conclusion, we expect this study to shed light on the impact of the
overlay structure on \bit{} efficiency, and to foster further
researches in that direction.

\subsection{Future Work}

Our future work will progress along two avenues.

\begin{itemize}

\item Mitigate the impact of NATed peers on the robustness
of the overlay. Actually, with its current implementation,
BitTorrent produces an overlay where non-NATed peers
have a higher connectivity than NATed ones. As a result,
one can create partitions by attacking the non-NATed peers, which are the
most connected ones. One possible solution to this problem
is to allow NATed peers to initiate more connections
than non-NATed ones. For example, one can imagine that the tracker
reports the number of NATed peers to new peers so that they can weight
their maximum number of outgoing connections. Our goal is to still
have a highly connected graph, but without peers with significantly
more connections. The intuition behind this solution is that
the robustness of the overlay would improve. This solution, and in
particular how to weight the maximum number of outgoing connections,
will be subject to further investigation.

\item Extend peer exchange in order to still converge fast to the maximum
peer set size while maintaining a low diameter overlay.
Indeed, with the current implementation of peer exchange, peers converge fast to
their maximum peer set size, but only peers that are at the
tail of the overlay chain have rooms for incoming connections.
As a result, new arriving peers can only connect to the
tail of the overlay chain. We are investigating possible
solutions to this problem whose main goal is to add randomness in the
overlay generated with peer exchange.

One solution is to allow peers coming from the tracker to
preempt connections of peers discovered with peer exchange. For
example, assume that peer $P_i$ has reached its maximum
peer set size and amongst its neighbors, there is $P_j$ that it
discovered with peer exchange. Assume now that $P_k$
joins the torrent and receives the IP address of $P_i$ from
the tracker. If $P_k$ initiates a connection to $P_i$, $P_i$ will
accept this connection and drop its connection to $P_j$.

Another solution is to add randomness during the construction of the
overlay. For instance, instead of collecting a list of neighbors of
its neighbors, which creates locality in the graph construction, a
peer $P_i$ can ask neighbors to randomly selected peers. The rational
is to discover and create connections to peers that are far from $P_i$
in the overlay. The choice of the random function to discover those
peers is critical and currently under investigation.

\end{itemize} 
\section*{Acknowledgments}
We would like to thank Olivier Chalouhi, Alon Rohter, and Paul Gardner 
from Azureus Inc. for their information on peer exchange. 

\bibliographystyle{IEEEtran}
\bibliography{biblio}

\end{document}